\documentstyle[preprint,aps,epsf,floats]{revtex}
  \def \ra       {\rightarrow}

  \def \cima     {ClO$^-_2$-I$^-$-MA }
  \def \cdima    {ClO$_2$-I$_2$-MA }
  \def \clo      { {\rm ClO}_2  }
  \def \clom     { {{\rm ClO}_2}^-}
  \def \ma       { {\rm MA}     }
  \def \ii       { {\rm I}_2    }
  \def \im       { {\rm I}^-    }
  \def \hp       { {\rm H}^+    }
  \def \siiim    { {{\rm SI}_3}^-}

  \def \maform   {CH$_2$(COOH)$_2$ }
  \def \CLO      { [{\rm ClO}_2] }
  \def \CLOM     { [{{\rm ClO}_2}^-] }
  \def \MA       { [{\rm MA}] }
  \def \II       { [{\rm I}_2] }
  \def \IM       { [{\rm I}^-] }
  \def \HP       { [{\rm H}^+] }
  \def \S        { [{\rm S}] }
  \def \SIIIM    { [{{\rm SI}_3}^-]} 
   
\begin{document}

%%%%%%%%%%%%%%%%%%%%%%%%%%%%%%%%%%%%%%%%%%%%%%%%%%%%%%%%%%%%%%%%%%%
%%-- These are needed for preprint style to save paper.
\draft 
\tighten
%%\preprint{HEP/123-qed}
%%%%%%%%%%%%%%%%%%%%%%%%%%%%%%%%%%%%%%%%%%%%%%%%%%%%%%%%%%%%%%%%%%%%

\title{Turing Instability in a Boundary-fed System}

\author{S. Setayeshgar\footnote{Corresponding author, Present address:
Physics Department, Northeastern University, Email: simas@masto.physics.neu.edu,
Fax: 617 373 2943, Telephone: 617 373 2944} and M. C. Cross}

\address{
Condensed Matter Physics 114-36, California Institute of Technology,
Pasadena, CA 91125
}
\date{\today}
\maketitle

%%%%%%%%%%%%%%%%%%%%%%%%%%%%%%%%%%%%%%%%%%%%%%%%%%%%%%%%%%%%%%%%%%%

\begin{abstract}

The formation of localized structures 
in the chlorine dioxide-idodine-malonic acid (CDIMA) 
reaction-diffusion system is investigated numerically 
using a realistic model of this system.  
We analyze the one-dimensional patterns formed along the 
gradients imposed by boundary feeds, and study their linear stability 
to symmetry-breaking perturbations 
(Turing instability) in the plane transverse to these gradients.  
We establish that an often-invoked 
simple local linear analysis which neglects
longitudinal diffusion is inappropriate for predicting the linear
stability of these patterns. 
Using a fully nonuniform analysis, 
we investigate the structure of the patterns formed along the 
gradients and their stability to transverse Turing pattern formation 
as a function of the values of two control parameters: the malonic acid 
feed concentration and the size of the reactor in the dimension along 
the gradients.  The results from this investigation are compared with 
existing experiments.  
\end{abstract}

\narrowtext

%%%%%%%%%%%%%%%%%%%%%%%%%%%%%%%%%%%%%%%%%%%%%%%%%%%%%%%%%%%%%%%%%%%%%%%%%

\section{Introduction}
\label{sec:intro}

Recent experimental developments have made possible the study of 
asymptotic spatiotemporal behavior in chemical systems in a
controlled and reproducible manner, allowing
predictions from theoretical and numerical studies of
these systems to be compared
quantitatively with experiments, in the same way that
fluid systems have been studied. Indeed, the
understanding of spatial
pattern formation in nonequilibrium systems has
greatly benefited from careful and controlled
experiments on fluid systems \cite{ref:CH_93}.
Unlike fluid systems
which at high nonlinearity break down to a turbulent state characterized by
a wide range of spatial scales, 
spatial patterns in chemical systems can
be studied at high nonlinearity \cite{ref:OuSw_95}, thus
providing an opportunity to
study rich and new phenomena that complement our
knowledge from pattern formation in fluid systems.  

The symmetry-breaking instability of a 
system from a homogeneous state to a patterned 
state, predicted in 1952 
by Alan Turing \cite{ref:Tur_52},  was 
observed for the first time nearly $40$ years later, in the
chlorite-iodide-malonic acid (CIMA) reaction-diffusion system
\cite{ref:CDBD_90,ref:DCDB_91,ref:OuSw_Nat91}.
The Turing instability is characterized by an intrinsic
wavelength resulting solely from reaction and diffusion 
processes.  For this reason, it has particular relevance to pattern formation 
in biological systems \cite{ref:Murray}.

In contrast to hydrodynamic systems for 
which the governing equations  and parameter values
are well understood, how to model complex
chemical systems is often not well known \cite{ref:CH_93}.  
A realistic model of the 
simpler chlorine dioxide-iodine-malonic acid (CDIMA)
reaction, which is similar to the CIMA reaction in terms
of its stationary pattern-forming and dynamical behavior,
has been proposed  by Lengyel, Rabai and Epstein
(henceforth referred to as  LRE) \cite{ref:LRE_90,ref:LRE2_90}.   
Hence, the CDIMA reaction-diffusion system 
has the potential to become an archetype for the
study of nonequilibrium pattern formation \cite{ref:DBDR_95}
in chemical systems, in principle allowing numerical and theoretical
investigations to be compared directly with experiments.

In practice, however, this has not been fully realized for two
reasons.
First, numerical investigations of reaction-diffusion equations using
realistic chemical parameters is a demanding computational
task. In addition, the algebraic complexity of the realistic nonlinear
reaction terms renders these models unsuitable for analysis by standard
analytical tools.
As a result, the theoretical
work on reaction-diffusion systems has been mostly based on
abstract models.
Second, despite the existence of the realistic CDIMA chemical
model which has similar pattern-forming and
dynamical properties to the related CIMA system,
experimental work has continued to be based on 
the CIMA reaction, making direct comparisons 
of numerical and analytical work with
experiments difficult.
Consequently, unlike in fluid systems, 
experimental and theoretical efforts
in chemical systems have not been closely coupled.

In this article, we use the realistic LRE model of the CDIMA
reaction-diffusion system to investigate
the Turing instability numerically \cite{ref:sth}.
Contrary to the case originally considered by Turing and
subsequently by others, the experimental conditions
under which Turing patterns form are not uniform,
as required by the continuous feed of reservoir
chemicals.   We study the formation and stability
of one-dimensional structures in the presence of boundary feed gradients.
We first briefly review the Turing mechanism in Section\ \ref{sec:turcond}.  
To facilitate comparisons with our numerical investigations, 
we describe the geometry and setup employed by the relevant experiments
in Section\ \ref{sec:exp}.
The LRE chemical 
model is described in Section\  \ref{sec:chemmod}.  In
Section\ \ref{sec:oned}, we obtain the one-dimensional 
steady state chemical concentration
profiles for a particular set of boundary conditions, and explore 
several different approaches to determine the
linear stability of these profiles to transverse symmetry-breaking patterns.
In Section\ \ref{sec:parsrch}, the patterns along the gradients and 
their linear stability are further
explored as a function of two control parameters.  We summarize our
results and consider prospects for further progress 
in Section\ \ref{sec:conclude}.

%%%%%%%%%%%%%%%%%%%%%%%%%%%%%%%%%%%%%%%%%%%%%%%%%%%%%%%%%%%%%%%%%%%%%%%

\section{Turing Instability Conditions}
\label{sec:turcond}

In his original paper \cite{ref:Tur_52}, Turing suggested that
the reaction and diffusion of chemicals
could account for the instability of an originally homogeneous steady 
state to a stable steady pattern when triggered by random disturbances.  
These instability conditions, which are derived and discussed in detail 
in the text by
Murray \cite{ref:Murray} are presented here for the purpose of introducing
the notation for the rest of the article.
Consider the general governing equations for the reaction-diffusion 
mechanism of two chemical species:
\begin{eqnarray}
\label{eq:gov1}
\sigma \frac{\partial u}{\partial t} & = & f(u,v; \vec{\mu}) + \nabla^2 u \\
\label{eq:gov2}
  \frac{\partial v}{\partial t} & = & g(u,v; \vec{\mu}) + c \nabla^2 v ,
\end{eqnarray}
where $f$ and $g$ represent the (nonlinear) reaction kinetics, $u$ and
$v$ are chemical concentrations, $\vec{\mu}$ is a set of reaction
parameters that may include concentrations of other chemical species,
$c=D_v/D_u$ is the ratio of diffusion
constants, and $\sigma \geq 1$ is a constant separating the
characteristic time scales for changes in the concentrations of the
$u$ and $v$ species.  Turing's idea was as follows \cite{ref:Murray}: if in 
the absence of diffusion $(u(\vec{r}, t), v(\vec{r}, t))$ 
tend to a linearly stable 
uniform steady state, then under certain conditions, 
the addition of diffusion leads to the development of
spatially inhomogeneous patterns.
Although these conditions were originally considered
for a spatially uniform system, where the parameters
$\vec{\mu}$ are constant, the actual experimental
realization of the Turing instability occurs
in the presence of externally imposed feed gradients,
where $\vec{\mu}=\vec{\mu}(z)$.  
In this section, we derive the Turing {\sl linear} instability conditions 
for both uniform and nonuniform parameters, $\vec{\mu}$.

%%%%%%%%%%%%%%%%%%%%%%%%%%%%%%%%%%%%%%%%%%%%%%%%%%%%%%%%%%%%%%%%%%%%%%%

\subsection{Uniform background}
\label{sec:unif}
The uniform background case is realized experimentally in batch
reactors where there are no externally imposed gradients from
continuous feed of chemical reactants, and Turing patterns
are necessarily transient.
The parameters $\vec{\mu}$ 
are constants independent of position.
The homogeneous steady state $\vec{c_s}=({u_s}^o, {v_s}^o)$ 
is obtained as the solution to:
\begin{equation}
\label{eq:uss}
  f(u,v;\vec{\mu}) = g(u,v;\vec{\mu}) = 0 .
\end{equation}
The linear stability of this state is obtained by substituting
into the governing reaction-diffusion equations:
\begin{eqnarray}
  \vec {c}(\vec {r},t) & = & \vec{c_s}
                           + \vec{\delta c}(\vec{r},t), \\
  \vec{\delta c}(\vec{r},t) & = & 
        \sum_k(\delta u_k^o, \delta v_k^o)\, 
        e^{i\vec{k}\cdot \vec{r}}\, e^{\lambda_k t}, 
\end{eqnarray}
where $\vec{c} = (u,v)$ is a vector of concentrations, $k$ is the spatial
wave number of the perturbation, $\lambda_k$ is the growth rate
of the $k^{\rm th}$ mode,
and $(\delta u_k^o, \delta v_k^o)$ is the corresponding constant
eigenvector.  The resulting linear eigenvalue problem:
\begin{equation}
 \label{eq:ie1}
 \left( \begin{array}{cc}
  \left( a_{11} - k^2 \right)/\sigma & a_{12}/\sigma \\
  a_{21}& a_{22} - c k^2  
      \end{array} \right)
      \left( \begin{array}{c}
           \delta u_k^o \\
           \delta v_k^o
      \end{array} \right)
      = \lambda_k
      \left( \begin{array}{c}
           \delta u_k^o \\
           \delta v_k^o
      \end{array} \right) 
\end{equation}
yields,
\begin{equation}
  \label{eq:l1}
  {\lambda_k}^{(\pm)} 
        = -   \frac{1}{2 \sigma} \left [(\sigma c+1)k^2 - 
                                         (a_{11}+\sigma a_{22}) \right ]
                  \pm \frac{1}{2 \sigma} \sqrt{
                  \left [(\sigma c+1)k^2 - (a_{11}+\sigma a_{22}) \right ]^2
                            - 4 h(k^2)},
\end{equation}
where
\begin{equation}
   h(k^2)=\sigma \left[  c k^4 - (a_{22} + c a_{11}) k^2 
                               + (a_{11} a_{22} - a_{12} a_{21}) \right].
\end{equation}
The quantities $a_{11} = f_u$, $a_{12} = f_v$, $a_{21} = g_u$, and 
$a_{22} = g_v$ are the elements of the Jacobian of the reaction
terms with respect to the concentrations, evaluated at the uniform 
steady state.  $\lambda_k$ has a rich behavior depending on the values of
$\sigma$, $c$, and $a _{ij}$.
The conditions for the Turing instability are that this uniform steady state
be linearly: (i) stable to homogeneous perturbation, and (ii) 
unstable to inhomogeneous perturbations. Hence, this is a 
{\sl symmetry-breaking} mechanism, 
since it breaks the homogenous spatial symmetry of the uniform state.
For the general reaction-diffusion system given in 
Eqs.\ (\ref{eq:gov1}) and (\ref{eq:gov2}), 
these conditions are derived in Appendix\ \ref{sec:app_cond}.
Below, we refer to the relevant results for the purpose of discussion.  
Stability of the uniform steady state to homogeneous $k=0$ perturbations
requires the following inequalities to be satisfied:
  \begin{equation}
  \label{eq:cond1}
           a_{11} + \sigma a_{22} < 0, 
  \end{equation}
  \begin{equation}
  \label{eq:cond2}
        a_{11}a_{22} - a_{12}a_{21} > 0 .
  \end{equation}
In order for the uniform steady state to be simultaneously unstable to
inhomogeneous $k \neq 0$ perturbations, we must have
  \begin{equation}
  \label{eq:cond3}
        a_{22} + c a_{11} > 0 ,
  \end{equation}
\begin{equation}
  \label{eq:cond4}
   (a_{22} + c a_{11})^2 - 4c (a_{11}a_{22} - a_{12}a_{21}) \geq 0.
\end{equation}
Comparing Eqs. (\ref{eq:cond1}) and (\ref{eq:cond3}) we conclude that $a_{11}$ 
and $a_{22}$ must have opposite sign.  In the standard terminology,
the activator species has a positive sign and the inhibitor has a 
negative sign in the Jacobian.
Thus, taking $a_{11}>0$ and $a_{22}<0$ identifies $u$ as the activator and 
$v$ as the inhibitor.  
If $\sigma=1$, then Eqs.\ (\ref{eq:cond1}) and \ (\ref{eq:cond3})
are simultaneously satisfied only for $c>1$.  In fact, given values of other
parameters, $c \gg 1$ is required.  Since diffusion
constants of ions in aqueous solutions are all nearly 
the same (${\cal{O}} (10^{-5}) {\rm{cm^2s^{-1}}}$), 
for the instability conditions to be satisfied, $\sigma$ must be
greater than one.
In Section \ \ref{sec:chemmod}, the requirement $\sigma >1$ will be
put in the context of the fortuitous role of the starch color
indicator in the pattern formation itself by providing the mechanism for
slowing the activator reaction and diffusion with respect to those of
the inhibitor.

Equations (\ref{eq:cond1})--(\ref{eq:cond4}) constitute the
Turing conditions.
It will be useful for future comparison of the local
stability analysis with the full stability analysis of the nonuniform steady
state to consider the Hopf bifurcation of the uniform system.
For reaction parameters such that
\begin{equation}
\label{eq:hopf_txt}
  (a_{11} - \sigma a_{22})^2 + 4\sigma a_{12} a_{21} < 0,
\end{equation}
there will be a complex conjugate pair of eigenvalues for
wave numbers in the range $0 < k^2 < {k_+^{({\rm H})}}^2$,
where ${k_+^{({\rm H})}}^2$ is given in Eq.\ (\ref{eq:kh_plus}).
With the above inequality satisfied, 
a Hopf bifurcation of the uniform system occurs 
when ($a_{11} + \sigma a_{22}) > 0$.
Beyond the Hopf bifurcation point, 
there will be an unstable complex conjugate pair
of eigenvalues for wave numbers in the range given 
by Eq.\ (\ref{eq:osc_unstab}).

%%%%%%%%%%%%%%%%%%%%%%%%%%%%%%%%%%%%%%%%%%%%%%%%%%%%%%%%%%%%%%%%%%%%%

\subsection{Nonuniform background}
\label{sec:nonunif}
In this case, the parameters $\vec{\mu}$, which depend on the concentrations 
of background chemicals fed through the boundaries, 
are not constant but rather are functions
of the variable $z$ along the direction perpendicular to the feed boundaries.
The steady state solution will now be a function of $z$, satisfying:
\begin{eqnarray}
  f(u_s(z),v_s(z); \vec{\mu}(z)) 
       + \frac{d^2 u_s}{dz^2} & = & 0, \label{eq:nonunifss1}\\
  g(u_s(z),v_s(z); \vec{\mu}(z)) 
       + c \frac{d^2 v_s}{dz^2} & = & 0, \label{eq:nonunifss2}
\end{eqnarray}
with Dirichlet boundary conditions at $z=0$ and $z=L_z$.
The stability of $(u_s(z),v_s(z))$ is given by linearizing
about this state:
\begin{eqnarray}
 \vec{c}(\vec{r},t) & = & \vec{c_s}(z)
              + \vec{\delta c}(\vec{r},t), \\
 \vec{\delta c}(\vec{r},t) & = & 
   \sum_{{k}_{\bot}} (\delta u_{{k}_{\bot}}(z), \delta v_{{k}_{\bot}}(z))\, 
   e^{i{\vec {k}}_{\bot}\cdot {\vec {r}}_{\bot}}\, e^{\lambda_{{k}_{\bot}} t},
\end{eqnarray}
where ${\vec{k}}_{\bot}$ is the wave vector perpendicular to the 
direction of the gradients.  
For simplicity of notation, we will drop the subscript ``$\bot$,'' 
taking $k$ to be the transverse wave number.
The resulting eigenvalue problem is: 
\begin{equation}
 \label{eq:ie2}
 \left( \begin{array}{cc}
  \left (a_{11}(z)+\frac{\partial^2}{\partial z^2} - k^2 \right )/\sigma & 
  a_{12}(z)/\sigma \\
  a_{21}(z) & a_{22}(z)+ c \frac{\partial^2}{\partial z^2} - c k^2  
      \end{array} \right)
      \left( \begin{array}{c}
           \delta u_k(z) \\
           \delta v_k(z) 
      \end{array} \right)
      = \lambda_k
      \left( \begin{array}{c}
           \delta u_k(z) \\
           \delta v_k(z) 
      \end{array} \right) ,
\end{equation}
with $(\delta u_k(z), \delta v_k(z))$ satisfying the same Dirichlet
boundary conditions as the steady state. 
As Pearson {\sl et al.} \cite{ref:PB_92} have noted, 
this is an infinite-dimensional
eigenvalue problem for each $k$ which is formally similar to the Schr\"odinger
equation.  However, the Jacobian of the reaction terms is not symmetric,
rendering the linear operator non-Hermitian.  Hence, 
it must be solved numerically, by discretizing the $z$ spatial direction
into $N_z$ mesh points and solving the resulting $2N_z \times 2N_z$ matrix
eigenvalue problem for each $k$.  This method of solution is described in
Section\ \ref{sec:linstab_nonuniform}.

%%%%%%%%%%%%%%%%%%%%%%%%%%%%%%%%%%%%%%%%%%%%%%%%%%%%%%%%%%%%%%%%%%%%%%%%%%%%%%

\subsection{Locally uniform background}
\label{sec:locallyuniform}
In the presence of ramps in control parameters, a naive assumption 
is that a structure will form in
the region of space where the local value of the control parameter allows it
to be stable in the corresponding uniform problem \cite{ref:PADD_92}.  
This ``locally uniform''
approach amounts to treating each location along the gradients in the 
$z$-direction to be an independent and uniform quasi-two-dimensional
system in the $x-y$ plane.  The
corresponding locally uniform steady state which depends parametrically
on $z$ is given by the solution to:
\begin{equation}
\label{eq:locallyuss}
  f(u,v;\vec{\mu}(z)) = g(u,v;\vec{\mu}(z)) = 0 .
\end{equation}
The Turing instability conditions can then
be examined at each point in $z$ to determine 
whether or not a linear analysis
predicts the formation of transverse Turing 
patterns in any interval along $z$. 

Since the resulting
eigenvalue problem for the stability of the locally homogeneous
steady state to a symmetry-breaking instability 
requires only a $2\times2$ analysis at each $z$, it is
computationally simple.  The validity of this local 
analysis is assessed in Sec.\ \ref{sec:linstab_uniform}, 
by comparing the result with that from the fully 
nonuniform analysis (Eq.\ \ref{eq:ie2})
of the steady state along the gradients.

%%%%%%%%%%%%%%%%%%%%%%%%%%%%%%%%%%%%%%%%%%%%%%%%%%%%%%%%%%%%%%%%%%%%%%%%%%%%%%%

\section{Experimental Geometry}
\label{sec:exp}

The first experimental realization 
of the steady-state Turing patterns predicted in 1952 
was made in 1990 by Castets {\it et al.}~\cite{ref:CDBD_90}, 
and was subsequently confirmed by others \cite{ref:DCDB_91,ref:OuSw_Nat91}. 
This was made possible by the development of open spatial reactors which
allowed experimentalists to maintain a reaction far from equilibrium through
a continuous supply of reactants, while avoiding convective transport.
These sustained patterns have been obtained in only one controlled
experimental system to date, the chlorite-iodide-malonic 
acid (CIMA) chemical reaction-diffusion system.  
The principles of operation of these reactors have been discussed elsewhere
in detail \cite{ref:CDBD_90,ref:DCDB_91,ref:OuSw_91,ref:DDRK_96}.

\begin{figure}
  \centering \leavevmode
  \epsfxsize=6.0in
  \epsfbox{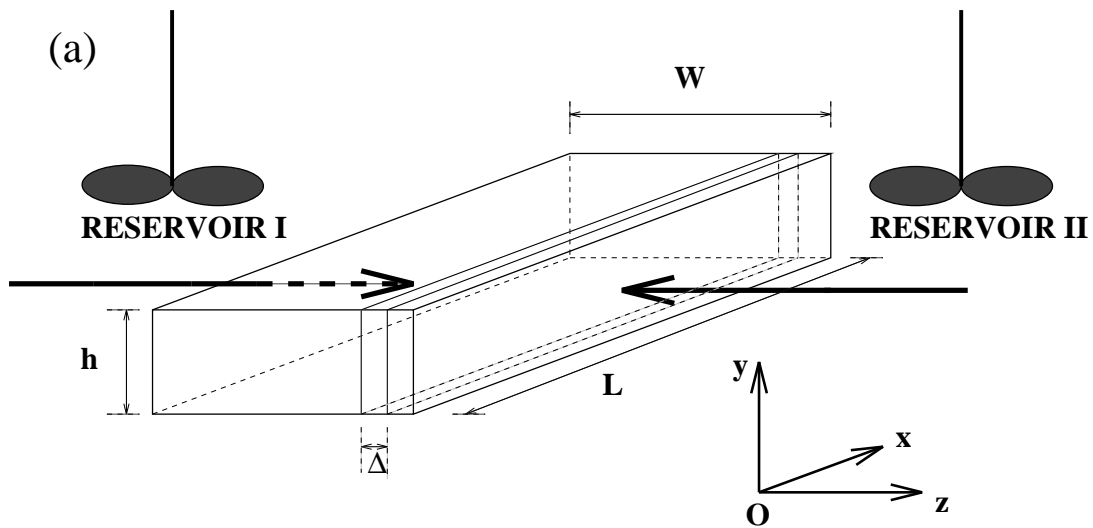}
  \caption[Schematic diagram showing typical experimental geometries]
           {\small
Sketch of open reactor geometries, adapted from 
Refs.~\cite{ref:DBDR_95,ref:DDRK_96}: 
A block of gel, with dimensions $L \gg w > h$ is in contact with 
two reservoirs I and II. The 
reservoirs are continuously stirred and fed with fresh supplies of 
reactants, such that each is separately nonreactive. 
A gradient in the reservoir species
forms perpendicular to the feed boundaries in the $z$-direction.  
The symmetry-breaking patterns form transverse to this gradient. 
The thickness $\Delta$ of the finite region along $z$ where they 
form is equal to at least one wavelength 
$\lambda$ of the Turing patterns.}
   \label{fig:fig1}
\end{figure}

In this section, we introduce the thin-strip reactor
which is investigated numerically in this work, since a geometrical
description of the experimental setup is useful in the 
understanding of our results.  
A detailed description of the chemical model is presented later
in Section\ \ref{sec:chemmod}, and is not necessary for 
the discussion presented here.  
The thin strip reactor is comprised of a thin rectangular gel strip, such 
that $L \gg w > h$, as in Figure\ \ref{fig:fig1}. Typically, $h < 1$ mm,
$L \sim 25$  mm, and $w \sim 3$ mm. The gels are water-based, acting as
essentially water in a loose polymer grid. The gel core of the 
reactor is in contact with two continuously stirred reservoirs 
of chemicals.  Components of the reaction are distributed in the 
two reservoirs in such a way that neither is 
separately reactive.  
In the CIMA experiments, these reservoir species are 
malonic acid (\maform or $\ma$), iodide ($\im$) and chlorite ($\clom$).
The LRE model of the CDIMA system takes as input
malonic acid, iodine ($\ii$) and chlorine dioxide ($\clo$).
As the reservoir species diffuse and react through 
the gel, the two dynamical species, iodide and chlorite,
which take part in the pattern-forming instability, are produced.  
The gel is preloaded with a soluble starch that acts as 
an indicator by changing color from yellow to purple with 
changes in triiodide concentration.  The large starch 
molecules are immobilized in the gel matrix, and for this reason 
actually play a role in the pattern formation itself.  

Observations are made in the direction 
perpendicular to the $x-z$ plane (along ${\rm O} y$), 
which allows viewing of the mulitfront patterns that develop 
along the boundary feed gradients (in the $z$-direction), 
as well as patterns that 
form parallel to the boundaries (in the $x$-direction), breaking
the boundary feed symmetry (single or multiple layers of spots). 
The symmetry-breaking patterns form in a
thickness $\Delta$ along $z$. If the gel is thin enough
($h \sim \lambda$ of the Turing patterns), these patterns
are one- or two-dimensional, depending on whether $\Delta$
is of the order of one or more wavelengths $\lambda$ of the structure.
With $h \gg \lambda$, for example $h \sim L$ as is the case
with disc reactors, the patterns are quasi-two-dimensional (referred to
as ``monolayers'') for $\Delta \sim \lambda$,  or three-dimensional
for $\Delta \geq \lambda$ (referred to as ``bilayers'' for 
$\Delta \sim 2 \lambda$).

A modified thin-strip reactor, where the feed surfaces ($L \times h$) are
no longer parallel but make an angle, has been developed and used by
Dulos {\sl et al.} \cite{ref:DDRK_96}, where
$h=0.2$ mm, $L=25$ mm, and the $w$ ranges from $1.75$ to $3.5$ mm. 
The variation in $w$ causes a gradual change 
in the reservoir concentration ramps across the gel, the effect of which
can be studied on the patterns that form along and transverse to the
gradients.

%%%%%%%%%%%%%%%%%%%%%%%%%%%%%%%%%%%%%%%%%%%%%%%%%%%%%%%%%%%%%%%%%%%%%%%%%%%%%%%

\section{Chemical Model of the CDIMA System}
\label{sec:chemmod}

Lengyel, Epstein and Rabai   
have proposed a model for the temporal oscillations
in the chlorite-iodide-malonic acid reaction, \cima, 
which is based on the simpler
chlorine dioxide-iodine-malonic acid reaction, \cdima, referred to as CDIMA
\cite{ref:LRE_90,ref:LRE2_90}.  They have shown 
experimentally that the CDIMA system also exhibits the 
Turing instability in both closed and open 
systems \cite{ref:LKE_92,ref:LKE_93}.
By monitoring the CIMA reaction in a closed 
system spectrophotometrically, it was 
determined that after an initial fast consumption of $\im$ and $\clom$
during a pre-oscillatory period to produce $\ii$ and $\clo$, 
the reaction of $\clo$, $\ii$ and $\ma$ accounts for the oscillations. 
The LRE CDIMA model consists of three reactions 
for $\ma$, $\ii$, $\clo$, $\im$, $\clom$ and $\hp$ \cite{ref:KLE_95},
with empirically determined rate laws:
\begin{eqnarray}
\label{eq:r1}
& &{\ma + \ii        \ra         {\rm I}\ma + \im + \hp} \nonumber \\ 
& &{-\frac{d\II}{dt}  =   \frac{k_{1a}\MA\II}{k_{1b} + \II}} 
     \equiv r_1        \\ [10pt]
\label{eq:r2}
& &{\clo + \im       \ra         \clom + 1/2~\ii}        \nonumber \\ 
& &{-\frac{d\CLO}{dt} =   k_2    \CLO \IM} 
     \equiv r_2                          \\ [10pt]
\label{eq:r3}
& &{\clom + 4\im + 4\hp  \ra    {\rm Cl}^- 
                            + 2\ii +2 {\rm H}_2{\rm 0}}  \nonumber \\ 
& &{-\frac{d\CLOM}{dt}  =  k_{3a} \CLOM \IM \HP+  
                          \frac{k_{3b} \CLOM \II \IM}{h+ \IM^2}} \equiv r_3. 
\end{eqnarray}

Lengyel {\sl et al.} \cite{ref:LE_92} have modeled the effect of 
unreactive starch-complex formation on the CDIMA system, where the complexing
agent is ($\rm S + \ii)$.  Although
formation of the starch-triiodide complex ($\siiim$) is a complicated process,
it can nevertheless be described as a single reaction:
\begin{equation}
\label{eq:r5}
 {\rm S} + \im + \ii \rightleftharpoons \siiim , \quad
         K = \frac{\SIIIM}{[{\rm S}]\IM\II} = \frac{k_+}{k_-},
\end{equation}
where $K$ is the equilibrium constant, and the reaction rate is given by:
\begin{equation}
  r_4 \equiv k_+ \S \II \IM - k_- \SIIIM.
\end{equation}
Using the above, the full reaction-diffusion model for the CDIMA system,
with the addition of the reaction with starch, is given by \cite{ref:KLE_95}:
\begin{eqnarray}
\label{eq:full_1}
\frac{\partial \MA}{\partial t}   &=& - r_1 + D_{\ma}\nabla^2\MA         \\
\label{eq:full_2}
\frac{\partial \II}{\partial t}   &=& - r_1 + \frac{1}{2} r_2 + 2 r_3 - r_4
                                            + D_{\ii}\nabla^2\II          \\
\label{eq:full_3}
\frac{\partial \CLO}{\partial t}  &=& - r_2 + D_{\clo}\nabla^2\CLO         \\
\label{eq:full_4}
\frac{\partial \IM}{\partial t}   &=&   r_1 - r_2 - 4 r_3 - r_4
                                            + D_{\im}\nabla^2\IM     \\
\label{eq:full_5}
\frac{\partial \CLOM}{\partial t} &=& r_2 - r_3
                                            + D_{\clom}\nabla^2\CLOM       \\
\label{eq:full_6}
\frac{\partial {\SIIIM}}{\partial t} &=& r_4                 \\
\label{eq:full_7}
\frac{\partial {\HP}}{\partial t} &=& r_1 - 4r_3 
                                      + D_{\hp}\nabla^2\HP. 
\end{eqnarray} 
The rate and diffusion constants
used in the numerical calculations here are taken from 
Refs.\ \cite{ref:KLE_95,ref:LE_91,ref:LE_95} and are given 
in Table \ref{table:T1}.

Lengyel {\sl et al.} \cite{ref:LRE_90} have shown that these 
reactions successfully simulate
the temporal behavior of $\MA$, $\II$, $\CLO$, $\IM$, and $\CLOM$
in a batch experimental system.
Their numerical results show that while the intermediates
$\IM$ and $\CLOM$ vary by several orders of magnitude during an oscillation,
$\MA$, $\II$ and $\CLO$ vary more slowly.
In addition, the contribution
of $\HP$ to the reaction terms is relatively small, and this species
can be neglected.
This suggests a reduction
of the full model to a three-variable system 
($\IM$, $\CLOM$ and $\SIIIM$) by treating the
concentrations of the $\MA$, $\II$ and $\CLO$
reactants as constants, making the model mathematically
and numerically more tractable. This procedure illustrates the
adiabatic elimination of fast modes in dynamical systems which reduces the 
full dynamics to only a few degrees of freedom. 

It is further assumed that there is a large excess of starch
uniformly distributed so that its concentration is always very
close to its initial value ${\S}_o$, and that the complex formation
and dissociation is fast.  Then,

\begin{equation}
\label{eq:s_approx}
\SIIIM\approx K {\S} {\II} \cdot \IM \approx K^\prime \IM,   
\end{equation}
where $K^\prime \equiv K {\S}_o {\II}_o$.
Adding Eqs.\ (\ref{eq:full_4}) and (\ref{eq:full_6}), and using 
Eq.\ (\ref{eq:s_approx}), a two-component reaction-diffusion system
is obtained:
\begin{eqnarray}
\label{eq:nr_1}
 \sigma \frac{\partial u}{\partial t} & = & {k_1}^\prime 
                - {k_2}^\prime u 
                - \frac{4 {k_{3b}}^\prime u v}{h + u^2} 
                + D_u \nabla^2 u     \\
\label{eq:nr_2}
\frac{\partial v}{\partial t} & = & {k_2}^\prime u                
                - \frac{  {k_{3b}}^\prime u v}{h + u^2}
                + D_v \nabla^2 v,
\end{eqnarray}
where ${k_1}^\prime = k_{1a} \MA_o \II_o/ ( k_{1b} + \II_o )$, 
${k_2}^\prime = k_2 \CLO_o$, ${k_{3b}}^\prime = k_{3b}\II_o$,
and $\sigma = 1 + K^\prime > 1$.  
The subscript ``o'' refers to the concentrations of species 
which are taken to be constant, and $u$ and $v$ represent the concentrations
of $\im$ and $\clom$ species.  
The role of the immobile starch color indicator in providing the relative
slowdown of the reaction and diffusion of the activator with
respect to that of the inhibitor enters through the parameter
$\sigma>1$.  

%%%%%%%%%%%%%%%%%%%%%%%%%%%%%%%%%%%%%%%%%%%%%%%%%%%%%%%%%%%%%%%%%%%%%%%

\section{Analysis}
\label{sec:analysis}

In this section, we use the LRE chemical model to 
investigate several aspects of the experimental
CIMA system. The focus is to demonstrate the
potential for quantitative analysis of experimental
results using the realistic CDIMA chemical model.
In Section\ \ref{sec:oned}, 
we investigate numerically the formation of one-dimensional
multifront localized structures along the gradients of imposed
boundary feeds. We study the linear stability of these structures to
transverse symmetry-breaking perturbations using the two-variable
reduction of the LRE model.  We compare our results from 
a local analysis to that from a
fully nonuniform analysis.  We review a proposed modification to the
local analysis and show that it does not successfully account for
the presence of gradients. In Section\ \ref{sec:parsrch}, 
we further explore the structure
and linear stability of the one-dimensional patterns along the boundary
feeds as a function of two control parameters: the malonic acid reservoir
concentration and gel width.  We map out the linear instability intervals
in each case.
We discuss the qualitative agreement of our results with relevant experiments.

%%%%%%%%%%%%%%%%%%%%%%%%%%%%%%%%%%%%%%%%%%%%%%%%%%%%%%%%%%%%%%%%%%%%%%%%%

\subsection{Linear Analysis of One-Dimensional Patterns Along Gradients}
\label{sec:oned}

\subsubsection{Stationary solution along the $z$-direction}
\label{sec:numres}

The full 7-component LRE model equations given in Eqs.\
(\ref{eq:full_1})--(\ref{eq:full_7}) were evolved forward in time to
obtain the steady state solution in one dimension along the gradients.
The boundary conditions,
${\MA}_{\small L}=1 \times 10^{-2}$ M at the left boundary, and 
${\II}_{\small R}=1 \times 10^{-3}$ M and
${\CLO}_{\small R}=6 \times 10^{-4}$ M at the right boundary,
were chosen so as to be consistent with a previous numerical investigation
of the LRE model in one dimension by Lengyel {\sl et al.} \cite{ref:LE_91}. 
Since the boundary conditions giving
a transverse instability were {\sl a priori} unknown, we used
these values as our starting point. 
The spatial $z$-direction was discretized on an 
irregularly spaced mesh to allow 
a greater number of mesh points in the regions where there was more
structure in the solution, without excessively increasing the overall number 
of mesh points in the problem.  A five-point finite-difference approximation
to the diffusion operator was used on the variable mesh.
The numerical scheme employed for the time evolution was Crank-Nicholson
implicit time-stepping for the linear terms, and Adams-Bashford explicit
time-stepping for the nonlinear terms. A banded solver \cite{ref:Num_Recp}
was used at each time step to solve for the solution at the next time step.  
The initial concentrations were uniform in the $z$-direction (and equal
to $5 \times 10^{-13}$ M). The time evolution was
continued until there was no appreciable change in the solution.  

\begin{figure}
   \centering \leavevmode
   \epsfysize=6.0in
   \epsfbox{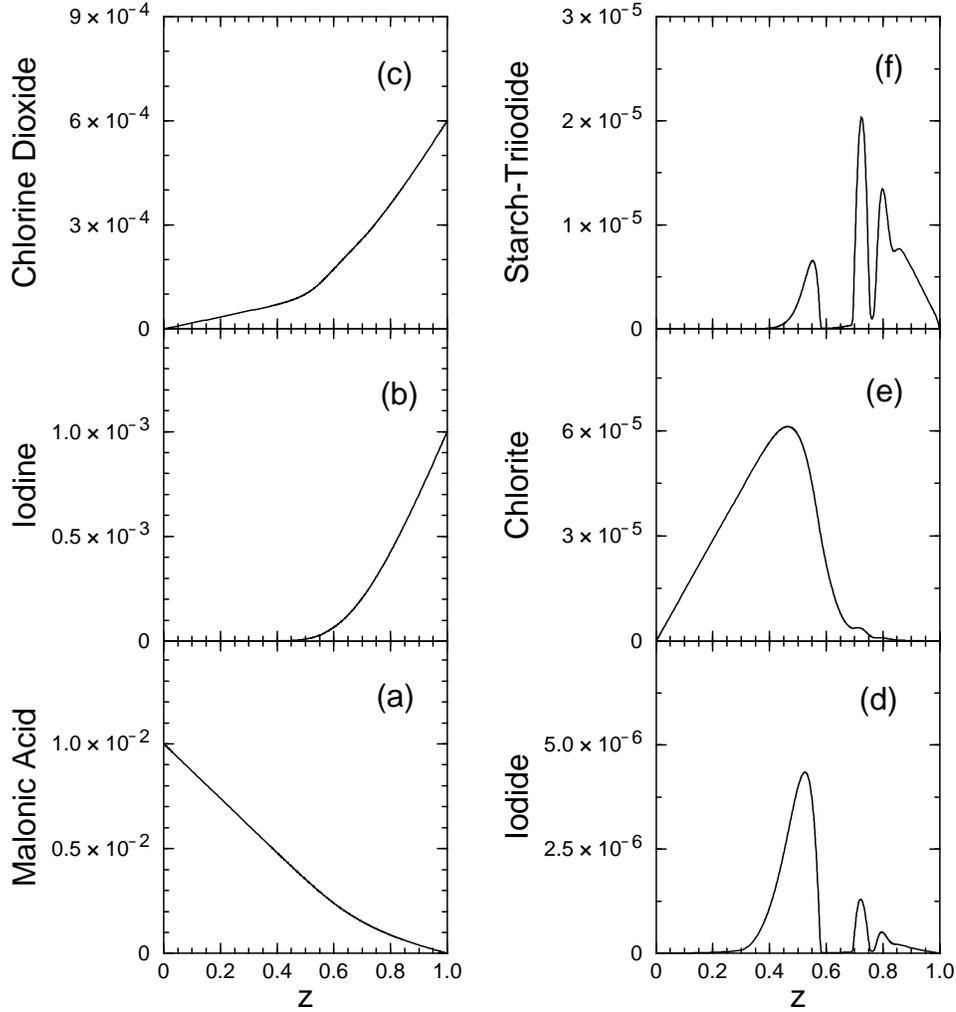}
   \caption[Steady state solutions for the 6-variable LRE model in one
            dimension]{
%\begin{sloppypar}
One-dimensional steady state solutions: The boundary conditions are 
${\MA}_{\small L}=1 \times 10^{-2}$ M at the left boundary, and 
${\II}_{\small R}=1 \times 10^{-3}$ M and
${\CLO}_{\small R}=6 \times 10^{-4}$ M at the right boundary.  
All other boundary conditions
are zero.  The $z$-axis has been normalized with respect to the thickness
of the gel in the $z$-direction, $w=0.3$ cm.  We note that the steady state
profiles for the reservoir variables malonic acid ($\ma$), iodine ($\ii$) 
and chlorine dioxide ($\clo$) vary considerably
from diffusion-only linear profiles.  The series of peaks in the 
starch triiodide ($\siiim$) profile correspond to experimentally 
observed stripes parallel to the feed boundaries \cite{ref:PADD_92}.
%\end{sloppypar}
}
   \label{fig:fig2}
\end{figure}

The results are displayed in Figure\ \ref{fig:fig2}. 
The steady state solution for the starch-triiodide complex
($\siiim$) plotted in Figure\ \ref{fig:fig2}(f)
represents the experimentally observable profile. 
As expected, it tracks the iodide ($\im$) 
steady state solution in Figure\ \ref{fig:fig2}(d). It corresponds to 
a (non-symmetry breaking)
pattern of stripes parallel to the feed boundaries 
(in the $x-z$ plane, with $x$ being the uniform direction),
such as those observed by Perraud {\sl et al.} 
\cite{ref:PADD_92}, although the
boundary species are different in their experiments on the CIMA reaction 
from those considered here. Upon increasing
the $\CLOM$ reservoir concentration, 
they observe a break-up of the stripes to rows of
spots parallel to the feed surfaces. This is a symmetry-breaking instability,
since the boundary feed symmetry of the system is broken.
In the following, we will investigate the linear stability of our
numerical steady state along the gradients to such a 
transverse pattern-forming instability.

%%%%%%%%%%%%%%%%%%%%%%%%%%%%%%%%%%%%%%%%%%%%%%%%%%%%%%%%%%%%%%%%%%%%%%%%%%%%%%%

\subsubsection{Locally uniform stability analysis}
\label{sec:linstab_uniform}
To examine the stability of the stationary patterns that
form along the gradients of boundary feeds ($z$-direction), the simplest
approach is to treat each location $z$ as being 
independent and locally uniform in the transverse plane
(see Section\ \ref{sec:locallyuniform}), 
thereby neglecting diffusion along the 
$z$-direction (longitudinal diffusion).
The locally homogeneous stationary state in the dynamical species,
$\im$ and $\clom$, at each $z$ can be constructed
either from the linear (diffusion only) profiles of the reservoir
species, $\ma$, $\ii$, and $\clo$, or more correctly, from their 
reaction$+$diffusion profiles, obtained
by evolving the full model, Eq.\ (\ref{eq:full_1})--(\ref{eq:full_7}).  
In either case, 
using the two-variable activator-inhibitor reduction
of the LRE model, Eqs.\ (\ref{eq:nr_1})--(\ref{eq:nr_2}), the resulting
eigenvalue problem for the stability of the locally homogeneous
steady state to a symmetry-breaking instability requires a simple $2\times2$
analysis. Hence, it is desirable to use such a local
approach if it can be shown that it accurately describes
the physical problem.  In that case, a transverse instability
would occur in a region of width $\Delta$ along the
gradients which is linearly unstable to a $k \neq 0$ 
instability. Indeed, even though Turing patterns are obtained 
under experimental conditions that by necessity
lead to nonhomogeneous parameter ramps, 
a local linear analysis is most commonly used to predict the
formation of a transverse symmetry-breaking instability.
In this section, we examine in the context of the
two-variable LRE model the locally uniform approach to determining the
stability of the stationary patterns that form along the
gradients of boundary feeds. 

\begin{figure}
   \centering \leavevmode
   \epsfysize=6.0in
   \epsfbox{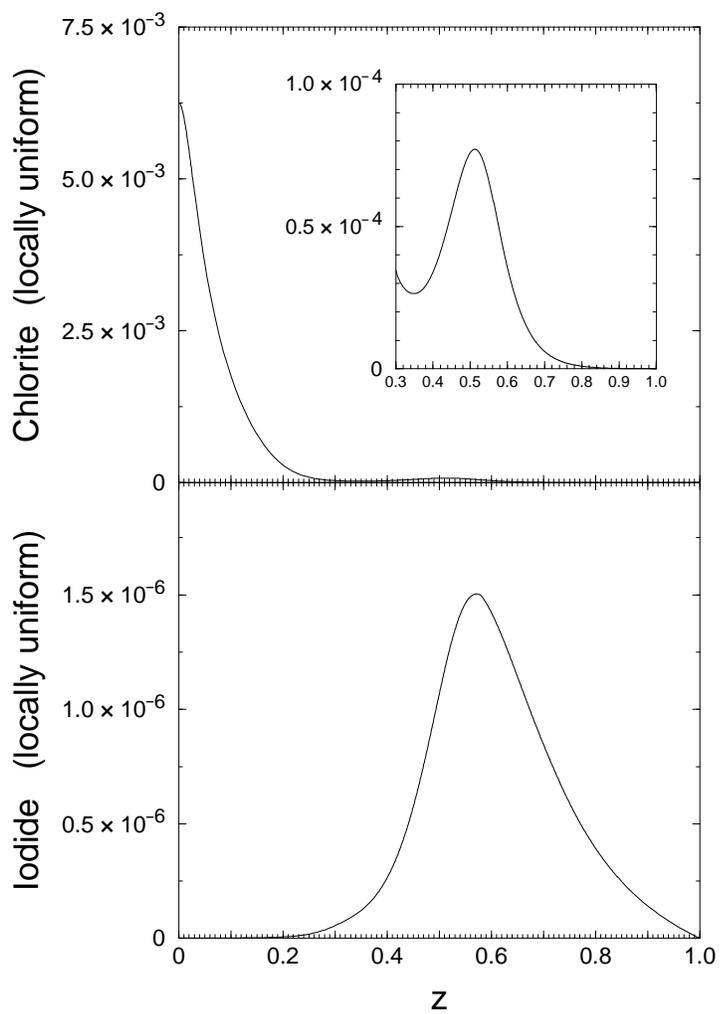}
   \caption[Locally uniform steady state solution for $\im$ and $\clom$]
{The value of the steady 
state solution, locally uniform in the $x-y$ plane, 
is plotted at each point $z$
for the two dynamical variables iodide ($\im$)
and chlorite ($\clom$) using the reaction $+$ diffusion 
profiles of the reservoir variables 
given in Figure\ \ref{fig:fig2}.}
   \label{fig:fig3}
\end{figure}

The locally uniform steady state in the variables
$\im$ and $\clom$ at each point in
$z$ along the gradients of the background chemicals is shown in
Figure\ \ref{fig:fig3}.
This solution is obtained according to 
Eqs.\ (\ref{eq:nr_1})--(\ref{eq:nr_2})
using the numerical reaction$+$diffusion profiles 
of Figure\ \ref{fig:fig2} for the $\ma$, $\ii$ and $\clo$ species, 
but neglecting the diffusion terms.
The dependence on $z$ in this plot is parametric.   

The stability of the local steady state at each $z$ is obtained from 
Eq.\ (\ref{eq:ie1}).
This analysis predicts the existence of a finite instability region.
The curves in Figure\ \ref{fig:fig4} represent the Turing
instability condition boundaries, 
Eqs.\ (\ref{eq:cond1})--(\ref{eq:cond4}),
for each locally uniform steady state.
We have also plotted the boundary which, when less than zero,
gives the interval in $z$ where the locally uniform steady state
has a complex conjugate pair of eigenvalues for a range in $k$ given 
by Eq.\ (\ref{eq:kh_plus}).  
The vertical axis has no labels 
since we have plotted quantities which have different dimensions 
and only their signs are of interest.   
The shaded region denotes the interval over which the
locally uniform steady state in the $x-y$ plane is linearly stable to 
homogeneous perturbations and unstable to inhomogeneous perturbations.
The width $\Delta$ of this region is approximately $0.15$ mm, which
is within the $0.13 - 0.33$ mm range of experimentally 
observed Turing wavelengths \cite{ref:OS_91}.
%Thus, the local linar stability analysis predicts that for the present control
%parameters, the experimental pattern of stripes will undergo an instability to
%symmetry-breaking spots in a finite region along the gradients.  

Figure\ \ref{fig:fig5} shows the gain curves for the locally uniform
steady states at selected points along the $z$-axis, consistent with the 
above linear stability boundaries.
Note that Figs.\ \ref{fig:fig5}(d)--(i) illustrate 
the role of the complexing agent 
$\left ({\rm S} + \ii \right)$ in suppressing the oscillatory
instability, since the concentration of $\ii$ 
(and therefore the complexing strength)
sharply increases as the right boundary is approached.     

%%%%%%%%%%%%%%%%%%%%%%%%%%%%%%%%%%%%%%%%%%%%%%%%%%%%%%%%%%%%%%%%%%%%%%%%%%%%%%%

\subsubsection{Fully nonuniform stability analysis}
\label{sec:linstab_nonuniform}

To assess the validity of the locally uniform stability analysis
presented above, we have carried out a fully nonuniform analysis, as
described in Section\ \ref{sec:nonunif}.
The eigenvalue problem given in Eq.\ (\ref{eq:ie2}) was 
discretized on the same variable mesh as that on which the nonuniform 
steady state was obtained, and was solved 
for all eigenvalues and eigenvectors at each value of transverse wave number
$k$ using {\tt{EISPACK}} \cite{ref:eispack}.  Since we have a general real 
matrix, with no special features such as symmetry, the most general routines
were used.  Figure\ \ref{fig:fig6} shows the real part of successive
eigenvalues with largest real parts.  
This result shows the steady state to be stable to all 
transverse perturbations.
%It corresponds to the experimental pattern of stripes parallel to the
%boundary feeds at a value of the control parameter such that there is
%no break-up to symmetry-breaking spots.
The eigenvector with slowest decaying 
(real) growth rate at $k=81.6$ ${\rm cm}^{-1}$ 
is plotted in Figure\ \ref{fig:fig7}.
It is localized roughly in the region along $z$ where the locally uniform 
stability analysis predicts the corresponding uniform steady state to be 
unstable to a transverse instability.

Since we are generally interested in the most unstable mode, 
in this case, we checked the numerical validity of the 
gain curve for the slowest decaying 
eigenvector (top-most continuous curve in Figure\ \ref{fig:fig6}) 
against two different numerical methods. First, the linear system 
for $2N_z$ variables (eigenvector), 
where $N_z$ is the number of mesh points, was 
solved as a nonlinear root finding problem in ($2N_z + 1$) variables, 
including the eigenvalue.    
Second, starting with the eigenvalue and eigenvector based
on the previous two methods, inverse iteration was used to 
verify the results.  Both
checks agree with the results from {\tt{EISPACK}}.  

The convergence of the slowest decaying gain curve as a function of the 
variable mesh was also investigated. Starting with a particular 
distribution of mesh points, the
mesh size was successively halved, the corresponding steady state obtained, 
and the eigenvalues and eigenvectors found.  
Basically, the number and distribution of mesh points must be sufficient to 
well resolve both the structure of the steady state and its most unstable 
eigenvectors
for numerical convergence. All numerical calculations were performed
on an {\tt{IBM RS6000}} workstation, with the exception of 
eigenvalue/eigenvector 
determination using {\tt{EISPACK}} with greater than 
approximately $500$ mesh points, 
which was done on a {\tt{CRAY C90}}. 

This result contradicts that from the locally uniform analysis 
which predicts a linear instability for these reaction parameters and
boundary conditions. Below, we review a proposed modification to the 
locally uniform analysis, and determine whether it is sufficient to
bring the local analysis closer to the fully nonuniform one.

In the above analysis and in those discussed in Section\
\ref{sec:linstab_uniform}, we have used the two-variable reduction of the 
LRE model.
We have directly verified the two-variable reduction of the full
$7$-variable (including $\hp$) model by performing a $7$-variable 
linear stability analysis of the 
nonuniform stationary state using inverse
iteration.  By comparing the $7$-variable and $2$-variable linear
stability results,
we have implicitly verified the assumption that the reservoir species, 
$\ma$, $\ii$ and $\clo$, do not play a role in determining the pattern-forming
instability of the stationary states \cite{ref:sth}.

%%%%%%%%%%%%%%%%%%%%%%%%%%%%%%%%%%%%%%%%%%%%%%%%%%%%%%%%%%%%%%%%%%%%%%%%%%%%%%%

\subsubsection{Modified local stability analysis}

The locally uniform analysis neglects
diffusion along $z$ which couples quasi-two-dimensional uniform slices.
Lengyel, Kadar and Epstein (LKE) \cite{ref:LKE_92} have proposed a 
modification to 
account for this diffusion, assuming that diffusion along $z$ is relevant 
only on a length scale of the order of the Turing wavelength. 
The basic idea behind the LKE 
modification is simple.  In the presence of gradients
in the $z$-direction, the Turing unstable mode is ``split'' between its 
``longitudinal'' (along $z$) and ``transverse'' dependence:
\begin{equation}
k_c^2=k_z^2+k_{\perp}^2,
\end{equation}
where $k_c$ is the critical wave number in the (narrow) Turing unstable
region along $z$, depending only on the local values of reaction and
diffusion parameters.
A transverse instability can occur provided the width of this 
region is not smaller than a Turing wavelength.
 
This modified local analysis is used to better predict the region along
the gradients where a transverse instability occurs, and to obtain
more accurately parameter values for investigating (transient)
Turing patterns in batch reactors.
The mechanics of the modification consists of adding an approximate 
term to the 
governing equations for the diffusion of the steady state along $z$, 
which does not
alter the composition of locally uniform steady state but does 
affect its stability.
This approximation to the diffusion operator is given by:
\begin{equation}
\frac{\partial^2 u}{\partial z^2} \approx \frac{u(z-\frac{\Delta}{2})-2 u(z)
                               +u(z+\frac{\Delta}{2})}{(\frac{\Delta}{2})^2}
                                  \approx \frac{8(\bar{u} - u(z))}{\Delta^2},
\end{equation}
$\bar{u}$ is the average value of the locally uniform steady state on the
two sides of the region of width $\Delta$, which is characteristic of
the longitudinal variation of the steady state.   
The validity of this estimate relies on the 
smallness of this width. The reaction terms $f$ and $g$ are modified:
\begin{eqnarray}
f^\prime(u,v;\vec{\mu}(z))=f(u,v;\vec{\mu}(z))+
                              8D_u((\bar{u} - u(z))/\Delta^2, \\ 
g^\prime(u,v;\vec{\mu}(z))=g(u,v;\vec{\mu}(z))+
                              8D_v((\bar{v} - v(z))/\Delta^2, 
\end{eqnarray} 
and the Jacobian of the reaction terms, $a_{ij}$, in the 
linear stability analysis is modified accordingly.
The Turing conditions can be rewritten as:
\begin{equation}
K^\prime(z) > H_1(z) > H_2(z),
\end{equation}
where
\begin{eqnarray}
   H_1 & \equiv & -a_{11}/a_{22}-1,	\\
   H_2 & \equiv &  
  -\frac{a_{11}}{2\sqrt{c(a_{11}a_{22}-a_{12}a_{21})}-ca_{11}}-1. 
\end{eqnarray}
The range of the 
Turing instability is given by the crosspoints of $K^\prime$
and $H_1$ and of $H_1$ and $H_2$.
Since $a_{ij}$ depend on $\Delta$ which is a priori unknown, 
$H_1$ and $H_2$ are evaluated iteratively from an initial estimate 
for $\Delta$ until convergence is achieved.  

\begin{figure}
   \centering \leavevmode
   \epsfxsize=6.0in
   \epsfbox{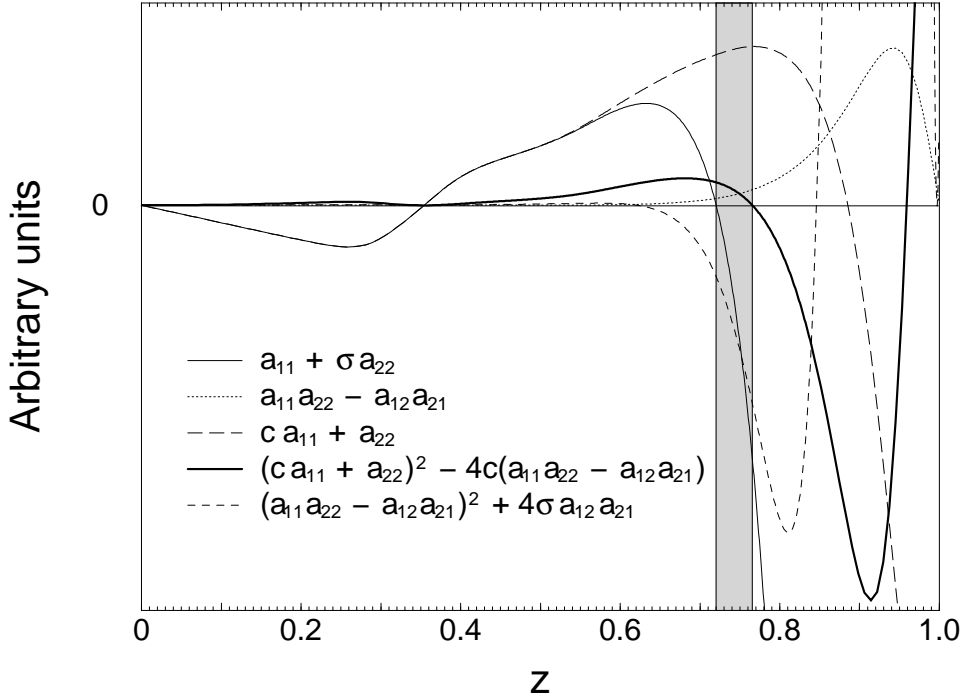}
   \caption[Boundaries from the locally uniform stability analysis]
{Boundaries from locally uniform stability analysis:  
At each location $z$, the
Turing instability conditions have been plotted for the corresponding uniform
steady state.  The plotted quantities have different dimensions, 
and since only the sign of each quantity is of interest, the vertical axis 
has no scale.  
When the light solid line corresponding to 
$\left (a_{11}+\sigma a_{22} \right)$ is less than zero, 
the complex conjugate 
pair of eigenvalues for $k=0$ has a negative real part. 
The dotted line is
$\left(a_{11}a_{22}-a_{12}a_{21}\right)$ which is additionally 
required to be greater 
than zero for stability of a real $k=0$ mode.  We note that this quantity
is everywhere greater than zero. 
The long-dashed line corresponds to
$\left(c a_{11}+a_{22}\right)$, and  
the heavy solid line is 
$\left(c a_{11}+a_{22})^2-4c(a_{11}a_{22}-a_{12}a_{21}\right)$, 
both of which must be greater than zero 
in order to have a $k \neq 0$ instability.  
The dashed line is 
$\left((a_{11}- \sigma a_{22})^2+4 \sigma a_{12}a_{21}\right)$, 
and where it is less than zero, 
a complex conjugate pair of eigenvalues exists for a
finite range in $k$.
The shaded region indicates where the locally uniform
steady state is stable to homogeneous perturbations and unstable to 
inhomogeneous perturbations. 
At $z=0.354$, the light solid $\left (a_{11}+\sigma a_{22} \right)$
and long-dashed $\left(c a_{11}+a_{22}\right)$ lines go through
zero, while the heavy solid, dotted and dashed lines remain positive.
}
   \label{fig:fig4}
\end{figure}

\begin{figure}
   \centering \leavevmode
   \epsfxsize=6.0in
   \epsfbox{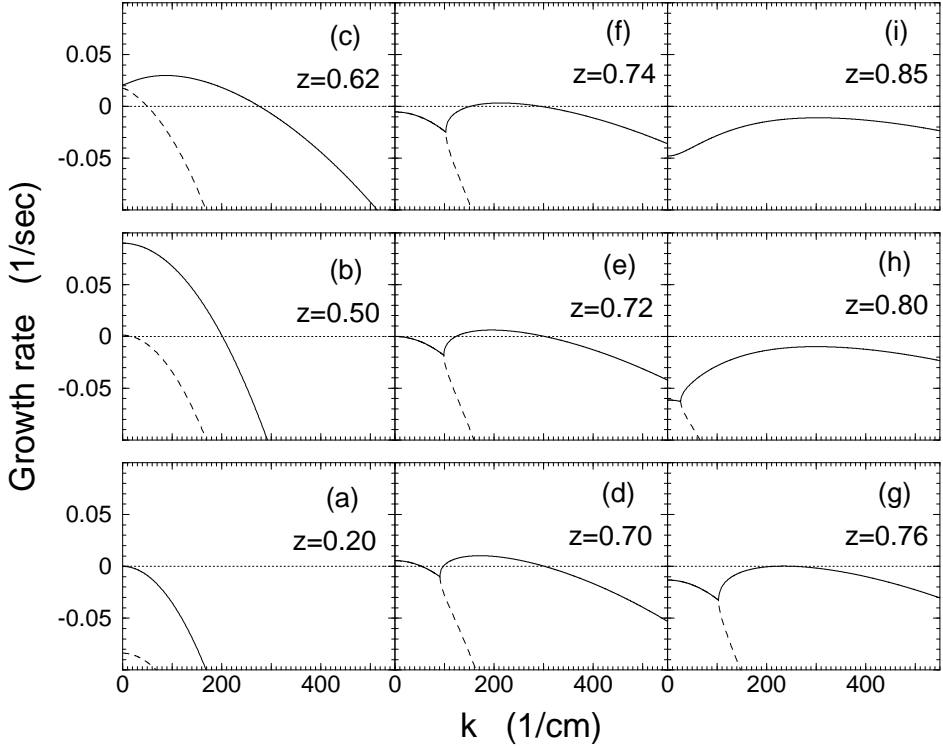}
   \caption[Gain curves for the locally uniform steady states]
{Gain curves for locally uniform steady states: The solid line denotes the 
eigenvalue with the larger real part, $\lambda_+$, and the dashed line 
denotes the one with the smaller real part, $\lambda_-$.
At (a) $z=0.20$, both eigenvalues are real and negative for all $k$;  
(b) $z=0.50$, both eigenvalues are real and peaked at $k=0$, with
$\lambda_+ > 0$ in the range $0<k^2<{k_+}^2$
and $\lambda_- > 0$ in the range $0<k^2<{k_-}^2$
(${k_{\pm}}^2$ are given in Appendix\ \ref{sec:app_cond}); 
(c) $z=0.62$, both eigenvalues are still real, but at $k=0$, we have 
a real degenerate pair, giving the boundary of the Hopf bifurcation;  
(d) $z=0.70$, complex conjugate pair of eigenvalues with positive
real part for $0<k^2<(a_{11}+\sigma a_{22})/(\sigma D_v+D_u)$,
and $k \neq 0$ instability for ${k_-}^2<k^2<{k_+}^2$;
(e) $z=0.72$, same as (d), except that the real part of the complex
conjugate pair is peaked at zero growth rate; 
(f) $z=0.74$, same as (d) with the real part of the complex conjugate pair 
less than zero at $k=0$;
(g) $z=0.76$, same as (f), except that the maximum growth rate for 
$k \neq 0$ is zero;
(h) $z=0.80$, same as (g), except that the maximum growth rate for
$k \neq 0$ is negative;
(i) $z=0.85$, both eigenvalues are real and negative.}

   \label{fig:fig5}
\end{figure}

\begin{figure}
   \centering \leavevmode
   \epsfxsize=6.0in
   \epsfbox{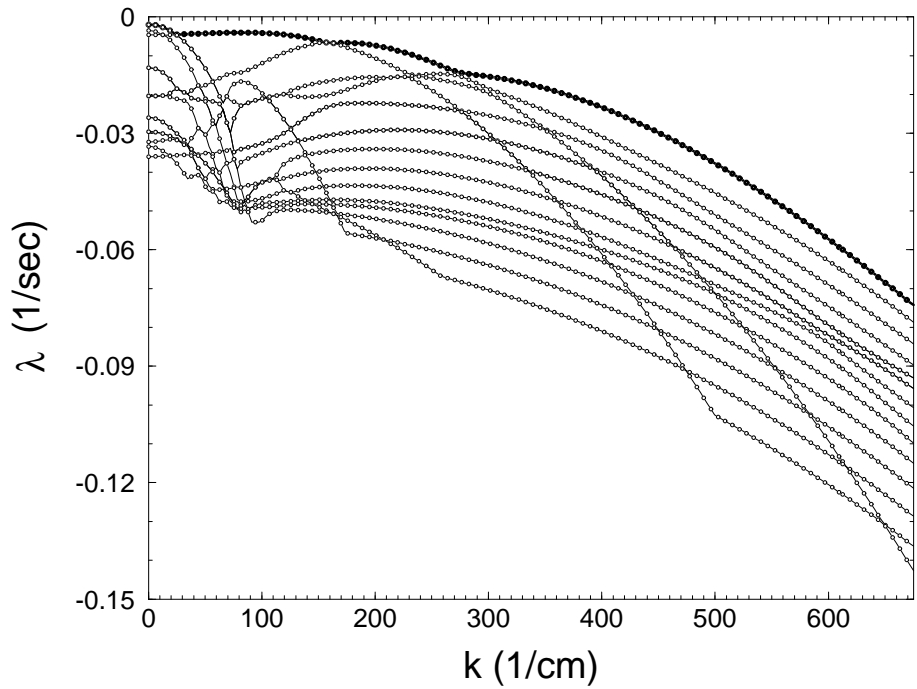}
   \caption[Gain curve for the one-dimensional steady 
            state along the gradients]
{Gain curve for the one-dimensional steady 
state along the gradients:  The
real parts of the first fourteen eigenvalues with largest real parts have
been plotted.  The eigenvalues are real, except at points (or along intervals)
where two curves intersect. 
The eigenvalue crossings appear imperfect due to the
coarse selection of $k$ values.  All eigenvalues are less than zero, with 
the heavy line corresponding to the slowest decaying mode at each $k$.}
%We
%note that the eigenvalue crossings define distinct ``modes'' that cut across 
%the spectrum of eigenvalues. This is further examined in 
%Figs.\ \ref{fig:fig14} and \ref{fig:mode}.}
   \label{fig:fig6}
\end{figure}

\begin{figure}
   \centering \leavevmode
   \epsfxsize=6.0in
   \epsfbox{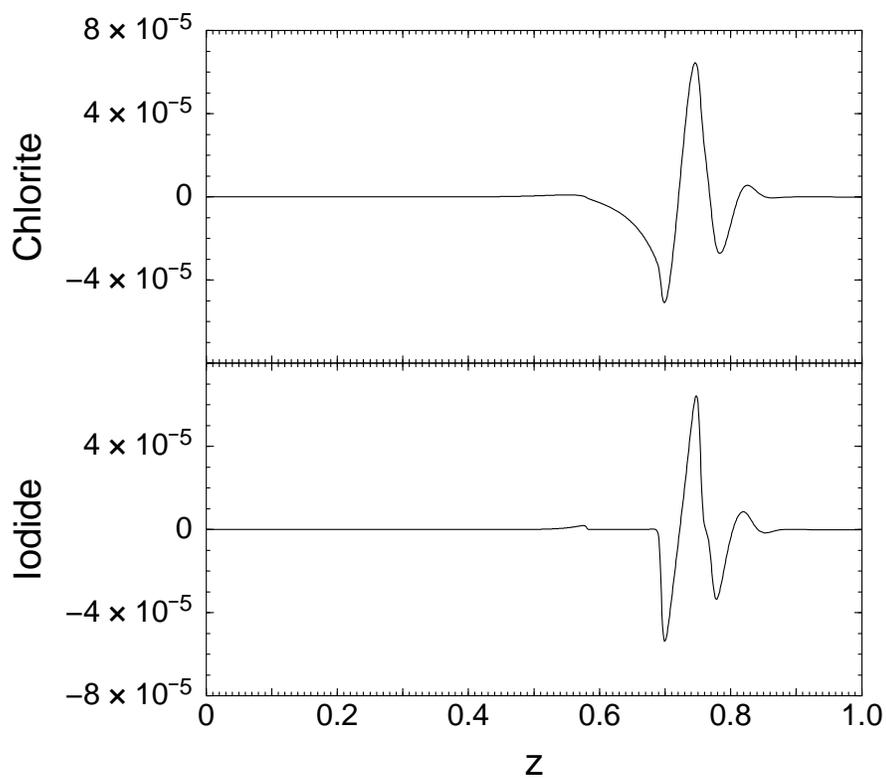}
   \caption[Eigenvector corresponding to the slowest decaying mode
in the fully nonuniform analysis]
{Eigenvector corresponding to slowest decaying mode in a
fully nonuniform analysis: The eigenvector at
$k=81.6\ {\rm cm}^{-1}$ with the largest real eigenvalue has been plotted.  
}
   \label{fig:fig7}
\end{figure}

\begin{figure}
   \centering \leavevmode
   \epsfxsize=6.0in
   \epsfbox{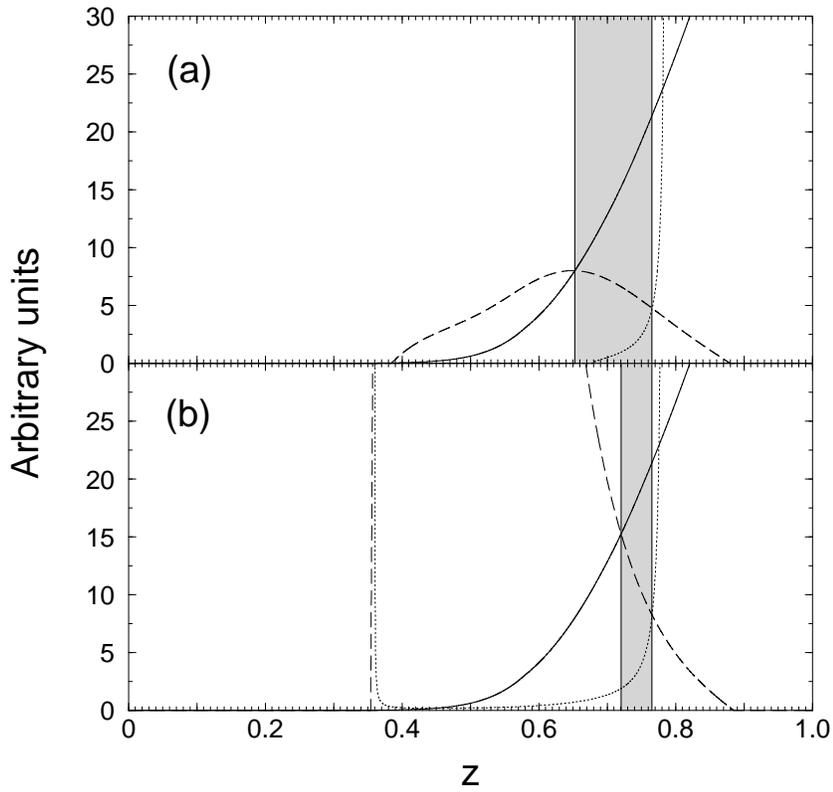}
   \caption[Boundaries in the modified local stability analysis]
{Modified local stability analysis: $K^\prime$, $H_1$ and $H_2$ are 
given by the solid, long-dashed and dotted lines respectively.
In (a), these boundaries
have been modified to take into account diffusion of the 
steady state along $z$,
whereas in (b) they are unchanged.
The Turing instability region is indicated by the shading.  The shaded
region in (b) is identical to that in Figure\ \ref{fig:fig4}.  Note
that the right boundary of the Turing region, which denotes $k \neq 0$ 
criticality, remains unchanged under the modification.  
}
   \label{fig:fig8}
\end{figure}

Figure\ \ref{fig:fig8} shows that the effect of this modification is to 
extend the $z$-range of the transverse instability from
approximately one to two Turing wavelengths.  The boundaries given by
the functions $K^\prime(z)$, $H_1(z)$ and $H_2(z)$ are combinations of the
boundaries given in Figure\ \ref{fig:fig4} which resulted directly
from the linear instability conditions.  Therefore, although the representation
of the boundaries in Figure\ \ref{fig:fig4} differs from that in
Figure\ \ref{fig:fig8}, 
the instability region is the same in the unmodified case.

The proposed modification increases the range of the Turing instability by
suppressing the homogeneous oscillatory 
instability (moving the left boundary
to the left), while not affecting the right boundary corresponding to
the inhomogeneous instability. Identifying the left boundary of the $z$-range
in which transverse patterns would form with the homogeneous instability
is unphysical, since there is no mixing of modes at the linear level. Instead,
it seems more appropriate to identify the left boundary of the Turing region
with the $z$-location where an inhomogeneous instability ceases to exist 
$\left(c a_{11} + a_{22} = 0 \right)$. 

An alternate modification to the local analysis is to 
carry out the linear stability
analysis about the nonuniform steady state along $z$, 
such that the diffusion operator in
the governing equations acts only on the steady state 
and not on the instability
eigenvector in the $z$-direction.  This ``local'' analysis  incorporates the
effect of diffusion along $z$ through the nonuniform steady state, but the
instability eigenvectors are ``local'' 
and depend only parametrically on $z$. The
resulting eigenvalue problem becomes:
\begin{equation}
 \label{eq:ie3}
 \left( \begin{array}{cc}
  \left (a_{11}(z) - k^2\right ) / \sigma & 
  a_{12}(z)/\sigma \\
  a_{21}(z) & a_{22}(z) - c k^2  
      \end{array} \right)
      \left( \begin{array}{c}
           \delta u_k(z) \\
           \delta v_k(z) 
      \end{array} \right)
      = \lambda_k(z)
      \left( \begin{array}{c}
           \delta u_k(z) \\
           \delta v_k(z) 
      \end{array} \right),
\end{equation}
where $a_{ij}(z)$ are evaluated at the nonuniform steady state, and
($\delta u_k(z), \delta v_k(z)$) and $\lambda_k(z)$ depend parametrically
on $z$. 
The stability boundaries are very irregular and not shown in this
case.  Except at the sharp edges of the nonuniform steady state
and over a region roughly equal to the width of the sharp edge
(much smaller than a Turing wavelength), this analysis 
predicts no $k \neq 0$ instability.

%%%%%%%%%%%%%%%%%%%%%%%%%%%%%%%%%%%%%%%%%%%%%%%%%%%%%%%%%%%%%%%%%%%%%%%

\subsubsection{Discussion}
\label{sec:conc} 
We conclude that to accurately predict the formation and location of
the Turing instability region, at least
the one-dimensional steady state along the gradients must be solved for
numerically using the full model including longitudinal diffusion.
A ``local'' stability analysis about this steady 
state does reproduce the result from the fully 
nonuniform stability analysis.  
%However, we have not shown that such a ``local'' stability analysis 
%based on the steady state which includes diffusion
%is successful in finding the instability region, since our choice of
%boundary conditions and reaction parameters has produced a completely
%stable state.
However, a local analysis neglecting longitudinal
diffusion of the stationary state does not correctly 
describe the linear stability of this state.
We have presented here a first direct demonstration 
of this point by carrying out a fully nonuniform as well
as a local linear stability analysis.
As has been suggested \cite{ref:LE_95}, two-dimensional (nonlinear)
time evolution of the model 
is the definitive method for predicting a
transverse instability.
We have accomplished this for the LRE model, and the results 
will be published elsewhere \cite{ref:ssmcc2}.
 
The locally uniform steady state
profiles for the two dynamical variables $\im$ and $\clom$ 
(Figure\ \ref{fig:fig3})
do not include diffusion along the $z$ direction 
and are qualitatively different from the numerical solution including 
diffusion (Figs.\ \ref{fig:fig2}(d), (e)). 
Hence, it is not surprising
that the local stability analysis about this steady state 
does not agree with the fully nonuniform one. 
In particular, at the left boundary, the locally uniform 
$\clom$ profile is several orders of magnitude greater than the corresponding
numerical solution including diffusion. 
This large discrepancy is accounted for by the diffusion of this species in
a region extending over approximately the left half of the gel, as can be
seen from the almost linear (diffusive) profile for $\clom$ over this region
(Figure\ \ref{fig:fig2}(e)). 
The LKE modification to the locally uniform analysis, which corrects for
this diffusion of the steady state along $z$, assumes that it
is relevant only on the length scale of the order of the Turing wavelength.
For the parameter values investigated here, this assumption is not valid,
and the modified stability analysis does not correctly predict the stability
of the structures along the gradients.

The dependence of the locally uniform steady state profiles for the 
intermediate species  
on the background concentration gradients is through the reaction terms.  The
results presented here are for background profiles obtained from reaction
and diffusion of all six species.  
We have checked that these background stationary
profiles do not vary considerably from those 
obtained by setting the intermediate
species identically equal to zero.  
In this way, we rule out the possibility that
the diffusion of the intermediate species feeds back 
into the profiles of the background variables, thereby accounting for
the large discrepancy between the locally uniform steady state profiles of the
intermediate species and those including diffusion.  In particular, the large
left boundary value of the locally uniform $\clom$ 
species results from the strong
suppression of $\ii$ relative to $\clo$ at this boundary,
which becomes even more pronounced with backgrounds obtained from the 
intermediate species set identically equal to zero.

It is desirable to obtain semi-analytical solutions to the
stationary structures along the gradients, which could then
be used in (semi-analytical) linear stability analyses of these states.  
This has been done, for example, for the Brusselator model
in the presence of slow spatial gradients
using a WKB-like approach \cite{ref:AN_75,ref:DB_89}.
The localized structures along the gradients are obtained 
as marginally stable perturbations
to the locally uniform steady state. 
However, in the case presented here, it is not possible to
carry out a similar analysis. 
First, there is a large discrepancy at the boundaries
between the locally uniform solution 
and the desired solution including diffusion 
satisfying Dirichlet boundary conditions.  Even if the boundaries are ignored
and an approximate solution in the interior of the gel is sought, our numerical
results show that the steady state including 
diffusion is not a weakly nonlinear
perturbation to the locally uniform steady state.  
Therefore, seeking a correction
given by marginally unstable modes is not justified.
For small gel width (see Sec.\ \ref{sec:parsrch_varl}), where the
background concentration profiles are almost linear, 
such a WKB-like description relying on the slowness of
the parameter ramps can perhaps be sought.

Numerical calculations based on the two-variable LRE model with
uniform background have shown the transition to a symmetry-breaking
instability to be strongly subcritical \cite{ref:JPDB_93,ref:JPMDB_93}.
Although it is not clear how the range of parameters investigated in these
works compares with their local values in the actual ramped experimental
system or in our numerical example, these results imply that a {\sl linear}
stability analysis of the locally homogeneous steady states would
not predict the existence of a finite amplitude instability in
the subcritical regime.  The nature of the transition of
the fully nonuniform stationary structures along the ramps to
a transverse symmetry-breaking instability 
has not been determined yet.
Should this transition be supercritical, or even weakly subcritical, 
the fully nonuniform
linear stability analysis would well predict the formation
of quasi-one-dimensional symmetry-breaking spots in the
thin-strip experiments.  This is currently under investigation
\cite{ref:ssmcc2}.
    
%%%%%%%%%%%%%%%%%%%%%%%%%%%%%%%%%%%%%%%%%%%%%%%%%%%%%%%%%%%%%%%%%%%%%%%%%%%%%%%

\subsection{Parameter Dependence of the Turing Instability in the CDIMA System}
\label{sec:parsrch}
As discussed in the previous section, 
by adopting a locally-uniform approach, 
existing numerical studies of the stability of the stationary 
patterns along feed gradients have not fully taken into account 
the reaction$+$diffusion feed gradients.
Thus, the parameter range for the occurrence 
of a transverse symmetry-breaking
instability in a gradient system within the context of the 
LRE model is essentially unknown.  Hence, we
have undertaken a systematic search, using the concentration 
of one of the reservoir
species and the gel width as the control parameters.  
Variation of either of the control parameters changes the
nonlinear reaction$+$diffusion profile of the background
species, however they are not equivalent operations. In the following sections,
we present our numerical results and make connection with relevant experimental
work.

\subsubsection{Variable malonic acid boundary condition}
\label{sec:parsrch_varma}
The parameter search for the Turing instability in the CDIMA reaction-diffusion
system as a function of the malonic acid concentration at the 
left boundary was performed for $\MA_{\small L}$ 
in the range $0.004$ M to $0.035$ M.  The concentrations
$\II_{\small R}$ and $\CLO_{\small R}$ at the right boundary 
were held fixed at $0.008$ M
and $0.006$ M, respectively.  
These values were chosen so as to lie within the range 
of the initial concentrations of these species 
used in experiments on this system 
in batch reactors \cite{ref:KLE_95}, and therefore should also be
experimentally accessible in open reactors. 

First, we numerically obtained steady state solutions 
of the full $7$-variable governing equations
as a function of $\MA_{\small L}$, as described in Section\ \ref{sec:numres}.
The analysis described in Section \ref{sec:linstab_nonuniform} 
of the linear stability along the gradients to transverse
symmetry-breaking perturbations was repeated for each stationary state.  
This was performed using the  
reduction of the full LRE model to the 
two dynamical variables $\im$ and $\clom$.
The eigenvalue and eigenvector corresponding
to the fastest growing (or slowest decaying) 
mode at each value of the transverse
wave number $k$ were obtained numerically using inverse iteration, 
and confirmed for
select values of $k$ using {\tt{EISPACK}} \cite{ref:eispack}.  

\begin{figure}
   \centering \leavevmode
   \epsfxsize=6.0in
   \epsfbox{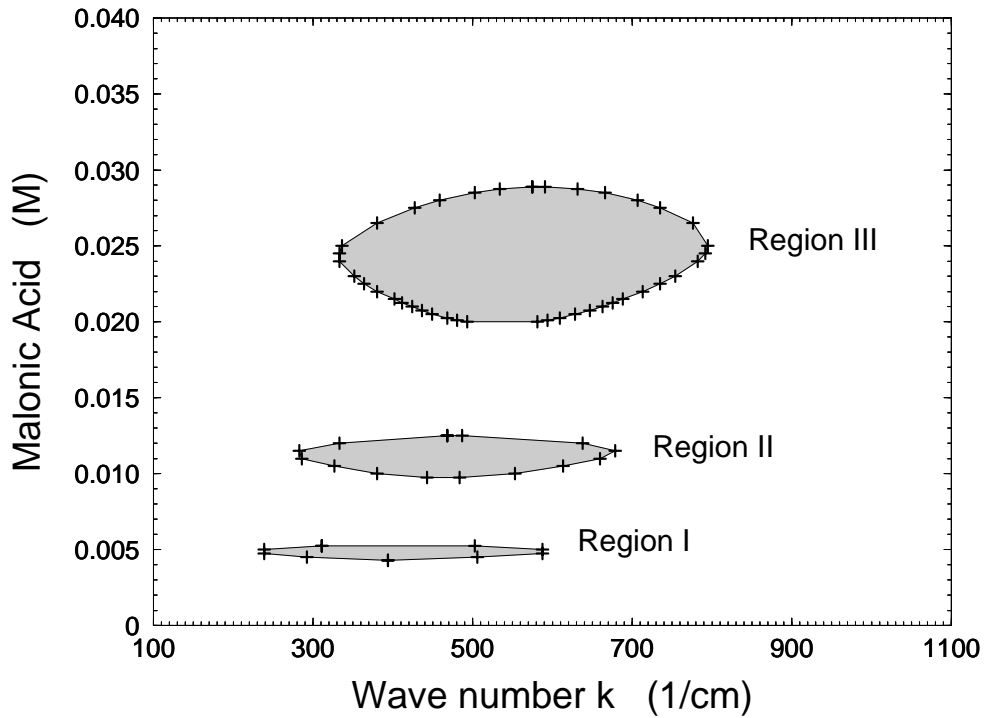}
   \vspace{0.2in}
   \caption[Malonic acid feed concentration at the 
left boundary versus the transverse
wave number]{\small
Malonic acid concentration (M) at the left boundary versus the 
wave number $k$ of the transverse instability (1/cm):  
The crosses represent the marginally stable wave numbers determined from
the stability analysis of the one-dimensional steady states corresponding
to the given malonic acid feed concentration. A transverse instability
occurs for values of malonic acid feed concentration in
the shaded regions, with the range of linearly unstable modes 
delimited by the solid lines for each value of $\MA_{\small L}$.  
}
   \label{fig:fig9}
\end{figure}

In Figure\ \ref{fig:fig9}, we have
plotted the value of the control parameter $\MA_{\small L}$ versus 
wave number, with the solid boundaries denoting marginally stable wave numbers.  
The shading indicates the Turing-unstable regions.  
We note that the unstable regions are disjoint, corresponding to the following
scenario: as the control parameter is 
continuously varied, the stable stationary state 
along the gradients first becomes unstable 
to transverse Turing patterns at a critical value of the control 
parameter, and initially remains
unstable as the control parameter continues 
to increase. It becomes stable again
once the control parameter exceeds a second and higher critical value.  
This is qualitatively 
consistent with the experimental observations of 
Perraud {\sl et al.} \cite{ref:PADD_92} 
on the CIMA reaction-diffusion system.   
Their results show that as the
concentration of $\CLOM_{\small R}$ at the right boundary is increased,
the number of alternating dark and bright bands parallel to the feed
boundaries increases, and several layers break up into rows of spots.
As $\CLOM_{\small R}$ continues to be increased,
the spot patterns develop along more bright bands, 
until they eventually disappear and the parallel stripes are recovered.

\begin{figure}
   \centering \leavevmode
   \epsfxsize=6.0in
   \epsfbox{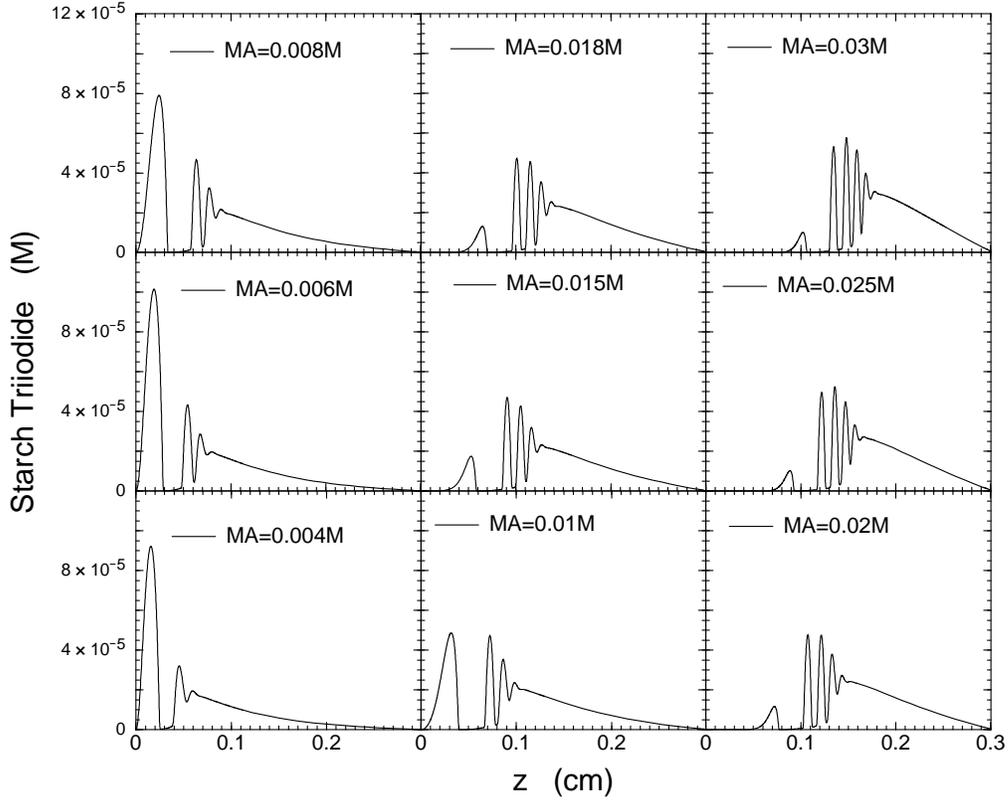}
   \caption[Stationary $\SIIIM$ solution along feed gradients 
for various values of malonic
acid feed concentration at the left boundary]{\small
Stationary solution for $\SIIIM$ along feed gradients 
for various values of malonic
acid feed concentration at the left boundary: 
This series of plots shows how the stationary state
along the gradients changes as $\MA_{\small L}$ is varied.
The number of peaks
in the solution increases with increasing malonic acid
concentration, while the left-most peak becomes smaller.
The minima correspond to light bands in the experimental
results.
Instability region I
corresponds to stationary states with three peaks, 
region II with four peaks and region III with five
peaks.
}
   \label{fig:fig10}
\end{figure}

\begin{figure}
   \centering \leavevmode
   \epsfxsize=6.0in
   \epsfbox{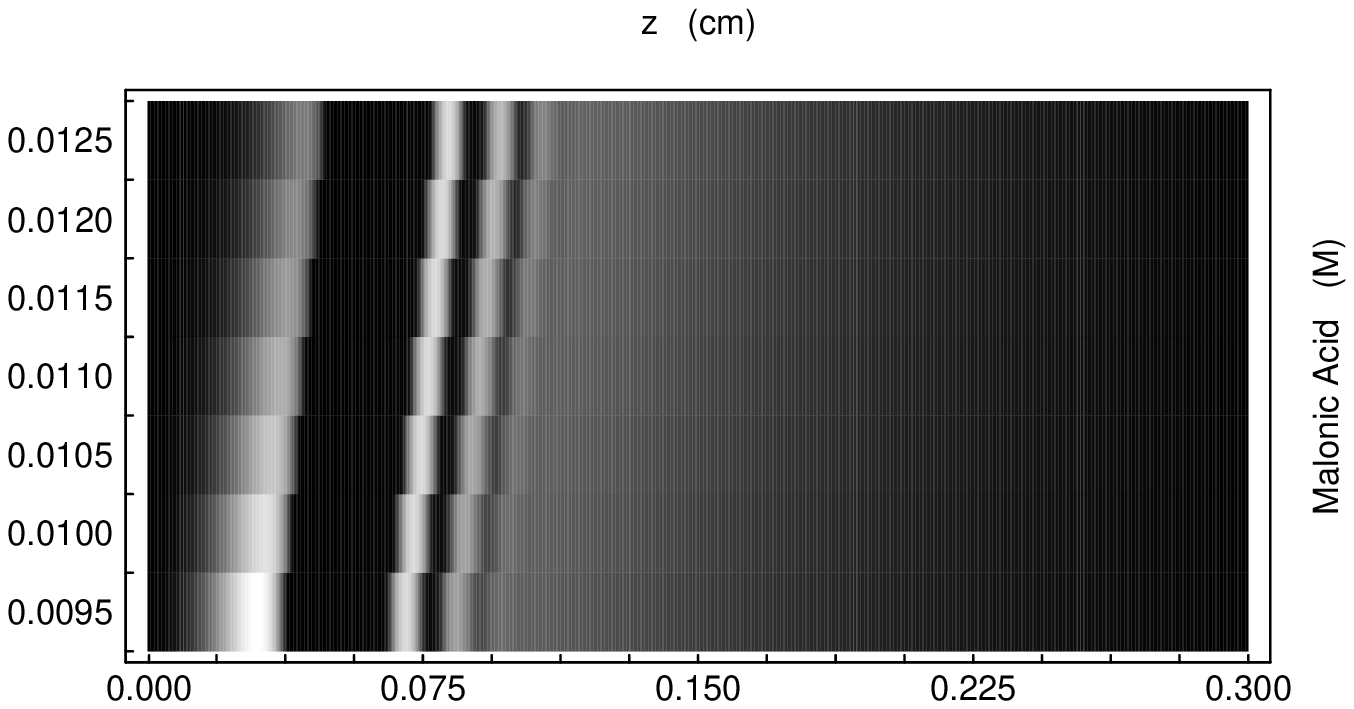}
   \vspace{0.2in}

   \centering \leavevmode
   \epsfxsize=6.0in
   \epsfbox{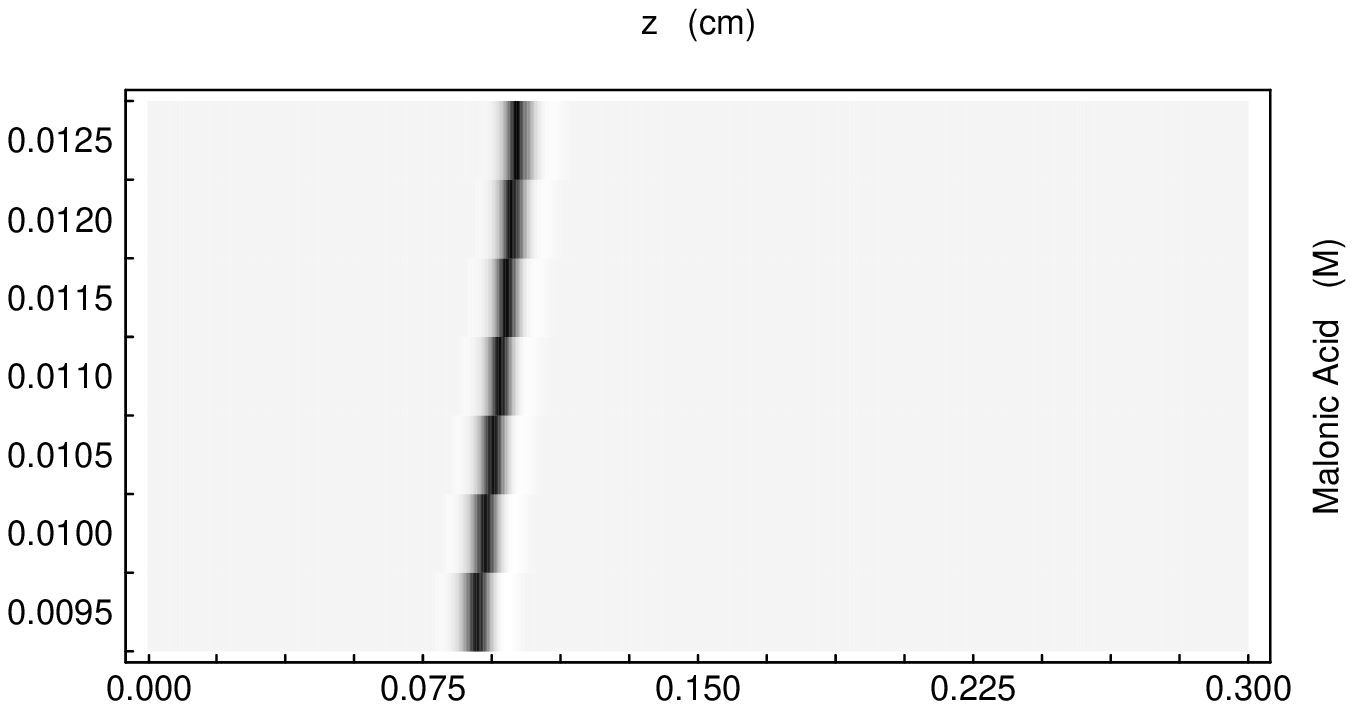}
   \vspace{0.2in}
   \caption[Density plots of the stationary $\SIIIM$ solution 
along the gradients and
the fastest-growing instability eigenvector for region II]{\small
Density plots of the stationary solution along the gradients and
the fastest-growing instability eigenvector for region II: White
and black correspond to large and small values of
the solution, respectively. The top
plot shows the variation of the stationary state in region II with
increasing $\MA_{\small L}$.  The bottom plot shows 
how the most linearly-unstable 
eigenvector is singly peaked and tracks the next-to-last minimum
in the solution. (Note:  It is difficult to discern the {\sl last} minimum 
in the top plot.) The ``staircase'' structure is due to the discrete
sampling of $\MA_{\small L}$.
}
   \label{fig:fig11}
\end{figure}

In Figure\ \ref{fig:fig10}, we have plotted the stationary solutions 
for the experimentally observed
$\siiim$ species from our numerical calculation, at various 
values of $\MA_{\small L}$. 
We observe that the number of peaks in the steady state 
solution increases with increasing $\MA_{\small L}$, while the characteristic 
``wavelength'' of the patterns along the gradients remains relatively constant
(excluding the leftmost peak).  The
transition from one unstable region to another corresponds roughly to the
appearance of an additional peak in the stationary solution along the gradient:
instability region I corresponds to a steady state with three peaks,
region II corresponds to four peaks, and region III corresponds to five peaks. 
As an example, for the instability region II,  
we show in Figure\ \ref{fig:fig11} density plots of the steady
state and the fastest growing (slowest decaying) wave vector.  
This eigenvector is localized at and roughly tracks one of the minima in 
the steady state solution. This is also the case for regions I and III.  
Specifically, the unstable vector appears to be approximately localized at 
the next-to-last minimum of the $\siiim$ solution.  
The appearance of the symmetry-breaking
instability at a minimum of the starch-triiodide is 
consistent with the above experimental observations.  

\begin{figure}
   \centering \leavevmode
   \epsfysize=6.0in
   \epsfbox{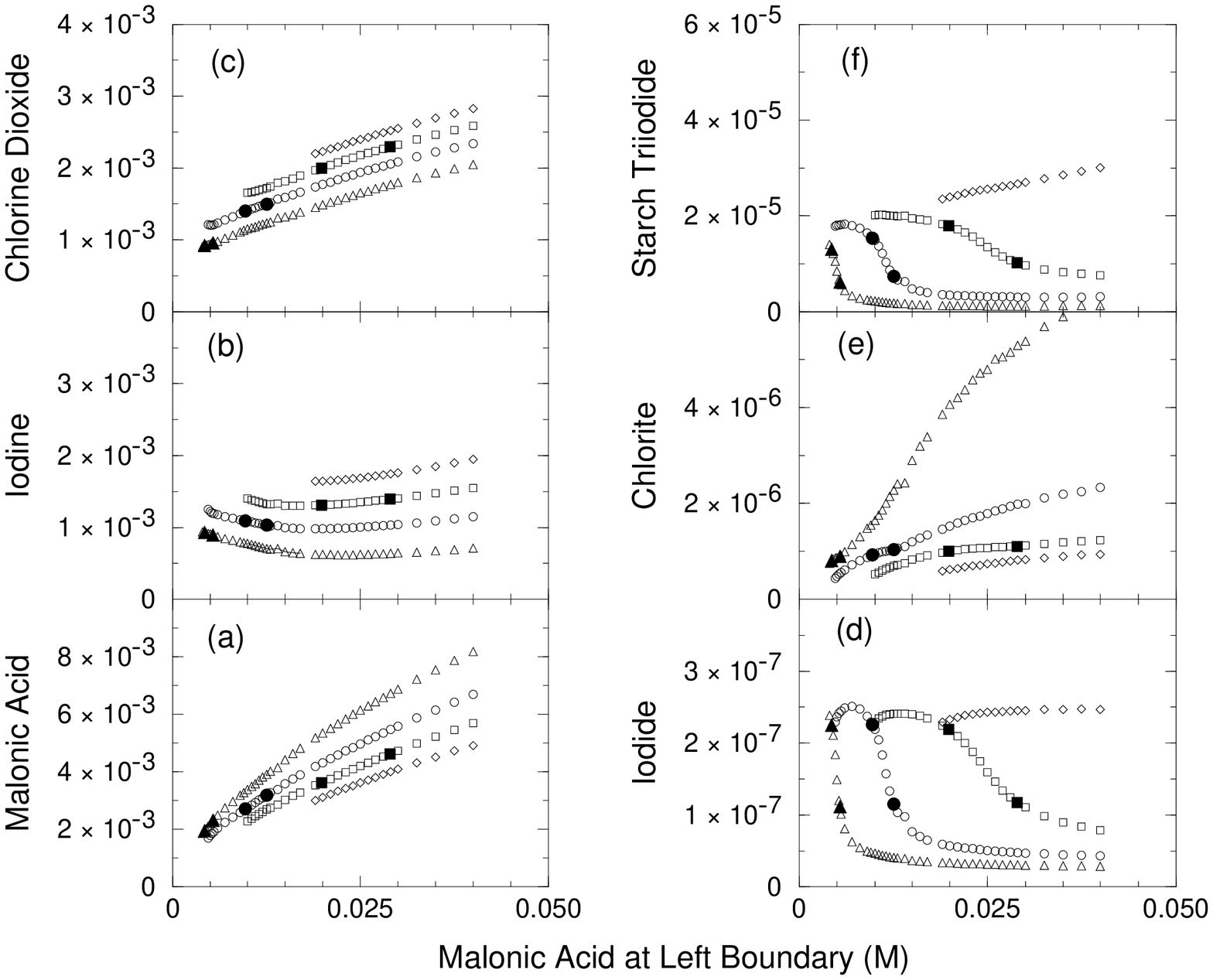}
   \vspace{0.2in}
   \caption[Values of stationary profiles of the 
CDIMA chemical species at
successive minima of the stationary $\SIIIM$ solution as
a function of $\MA_{\small L}$]{\small
Values of stationary profiles of the CDIMA chemical species at
successive minima of the stationary $\SIIIM$ solution as
a function of $\MA_{\small L}$:  The open triangles, circles, squares
and diamonds represent the second, third, fourth and fifth minimum,
respectively.  The filled symbols correspond to the critical
values of $\MA_{\small L}$; the points between the filled symbols
correspond to the linearly unstable states.  We note that the
instability occurs successively along the second, third and fourth minimum.
Although the values of $\MA$, $\II$, $\CLO$ in 
the unstable ranges continue
to increase as functions of $\MA_{\small L}$ in going from one instability
region to the next, the trend for the dynamical species $\IM$ and
$\CLOM$ is different: the value of $\CLOM$ remains approximately
constant, while $\IM$ varies within an approximately constant
range.
}
   \label{fig:fig12}
\end{figure}

To better quantify this trend, in Figure\ \ref{fig:fig12}, we plot the
value of all six chemical species at successive minima of the $\siiim$
stationary solution as a function of $\MA_{\small L}$.  
In this figure, the circles, triangles, squares and diamonds
correspond to the second, third, fourth and fifth minima, respectively.  
The points corresponding to critical $\MA_{\small L}$ values
are filled (two points for each unstable region). 
We note that the relevant chemical species for tracking the instability
are the two dynamical species, $\im$ and $\clom$, plotted in 
Figure\ \ref{fig:fig12}(d) and (e). 
(The $\siiim$ concentration depends on the product of the $\im$ and
$\ii$ concentrations.)  The concentration
of $\im$ remains within the range of approximately 
$\left(1.0-2.2\right)\times10^{-7}$M, and
the concentration of $\clom$ remains approximately constant at $1\times10^{-6}$
in each of the instability regions. (Note that the concentration of $\siiim$
does not stay within a more-or-less constant range for each of the instability
regions due to the drift in the $\ii$ concentration.)
A simple interpretation of these numerical results is that an 
instability occurs when the concentrations
of the dynamical species $\im$ and $\clom$ lie within a certain range.
As the stationary solution changes with increasing $\MA_{\small L}$, 
the instability vanishes when the value of the
concentration of the steady state falls outside of this range, and reappears
when the concentration at the {\sl next} minimum is within this range again.
  
\begin{figure}
   \centering \leavevmode
   \epsfxsize=6.0in
   \epsfbox{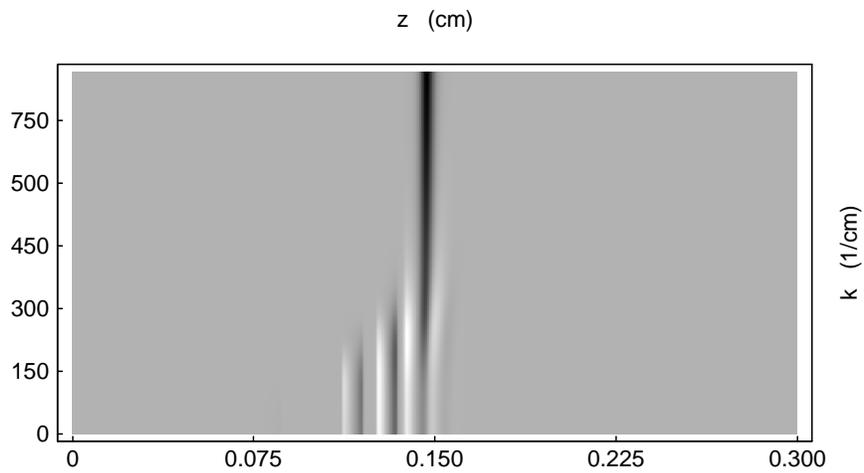}
   \vspace{0.2in}
   \caption[Density plot of the iodide eigenvector corresponding 
to the largest eigenvalue
as a function of the transverse wave number 
$k$ and spatial distance along
the gradients ($\MA_{\small L}=0.023$ M)]{\small
Density plot of the iodide eigenvector corresponding to the largest eigenvalue
as a function of the transverse wave number $k$ and spatial distance along
the gradients ($\MA_{\small L}=0.023$ M): Black and white correspond 
to low and high values of the
eigenfunction, respectively.  For small values of $k$, the eigenvector is
multiply peaked while for $k$ larger than approximately $300$ $\rm{cm^{-1}}$,
which includes the unstable range of wave numbers, it is singly peaked.
}
   \label{fig:fig13}
\end{figure}

The experiments
of Perraud {\sl et al.} \cite{ref:PADD_92} show that as the 
control parameter is increased, 
the stationary pattern of stripes along the
gradients first becomes unstable in
a single stripe region and then in multiple stripes as the control parameter 
continues to increase.  In our numerical investigations, the most unstable
vector remains singly peaked in all cases.  To further investigate this
point, we have examined the spatial structure (along $z$)
of the linear instability eigenvectors
as a function of $k$. 
Figure\ \ref{fig:fig13} shows a density plot of the most 
unstable (or least stable) eigenvector, corresponding to the $\im$ species, 
as a function of transverse wave number $k$ for $\MA_{\small L}=0.023$ M.
The horizontal
axis represents the spatial coordinate along the gradients, and the vertical
axis is the transverse wave number, ranging from $k=0$ $\rm{cm^{-1}}$ 
to $k=900$ $\rm{cm^{-1}}$.  

\begin{figure}
   \centering \leavevmode
   \epsfxsize=6.0in
   \epsfbox{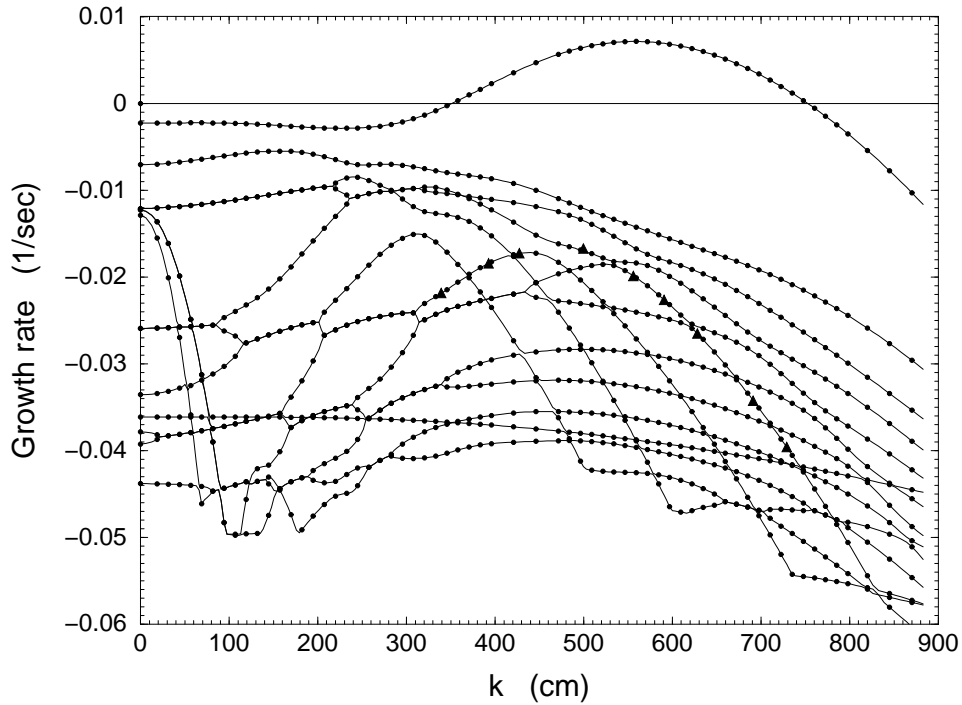}
   \vspace{0.2in}
   \caption[Spectrum of eigenvalues for $\MA_{\small L}=0.023$ M]{\small
Spectrum of eigenvalues for $\MA_{\small L}=0.023$ M: The real parts of the
first fourteen eigenvalues with largest real parts have been plotted.
The eigenvalues are real, except along intervals where two curves
overlap.  Some eigenvalue crossings appear imperfect due to the 
coarse selection of $k$ values.  It appears that the first two
eigenspectra do not cross but remain distinct. 
We note that the eigenvalue crossings define distinct ``modes'' that
cut across the spectrum of eigenvalues. 
The filled triangles indicate the eigenvalues for one such ``mode''.

}
   \label{fig:fig14}
\end{figure}

In Figure\ \ref{fig:fig14}, we show 
the first fourteen eigenvalues with largest
real parts as a function of $k$.
We note the eigenvalue crossings that define distinct ``modes''
cutting across the spectrum of eigenvalues. As an interesting aside,
we have investigated the eigenvectors corresponding to one such ``mode'',
denoted by filled triangles in this figure.
The spatial profile of these eigenvectors
changes continuously with increasing $k$.  Hence, their
interpretation as a single physical ``mode'' is not obvious.
There does not appear to be
a mode crossing between the 
first and second branches of Figure\ \ref{fig:fig14}, although 
the structure of the first eigenvector changes considerably from being
multipeaked (and stable) to being singly peaked (and unstable, but subsequently
stable again) as $k$ increases.
The experiments of Perraud {\sl et al.} could correspond
to a case where, as the control parameter is varied, a mode with multiple
peaks becomes unstable.  This does not occur in our numerical investigation,
where the multiply peaked modes remain stable. The appearance
of multiple instability layers could also be a nonlinear effect, resulting
from linear growth and nonlinear saturation of the 
singly peaked unstable eigenvector, as well as growth of its side lobes.
However,  we have shown \cite{ref:ssmcc2}  
that in the two-dimensional evolution
of the governing equations, the spatial profile of the most linearly
unstable mode is in fact preserved in the LRE model.  

%%%%%%%%%%%%%%%%%%%%%%%%%%%%%%%%%%%%%%%%%%%%%%%%%%%%%%%%%%%%%%%%%%%%%%%%%%%%%%

\subsubsection{Variable reservoir length along boundary feeds}
\label{sec:parsrch_varl}

The experiments of Dulos {\sl et al.} \cite{ref:DDRK_96} have aimed at 
elucidating the transition from
quasi-two-dimensional to three-dimensional Turing patterns
by combining observations in bevelled thin-strip and disc reactors.
Motivated by these experiments, 
we have undertaken
a similar numerical investigation which does not 
directly address the same question, but rather continues to focus
on the localized patterns along the ramps.  In particular, we consider the
experimental results from the variable-width thin-strip reactor.
In these experiments, the transition between the domains with
one and two rows of spots, and the possible influence of the feed gradients
on the phase relations between the spots in the two 
rows have been studied.  The vanishing
amplitude of the spot modulations before this splitting occurs is not
well understood.  

\begin{figure}
   \centering \leavevmode
   \epsfxsize=6.0in
   \epsfbox{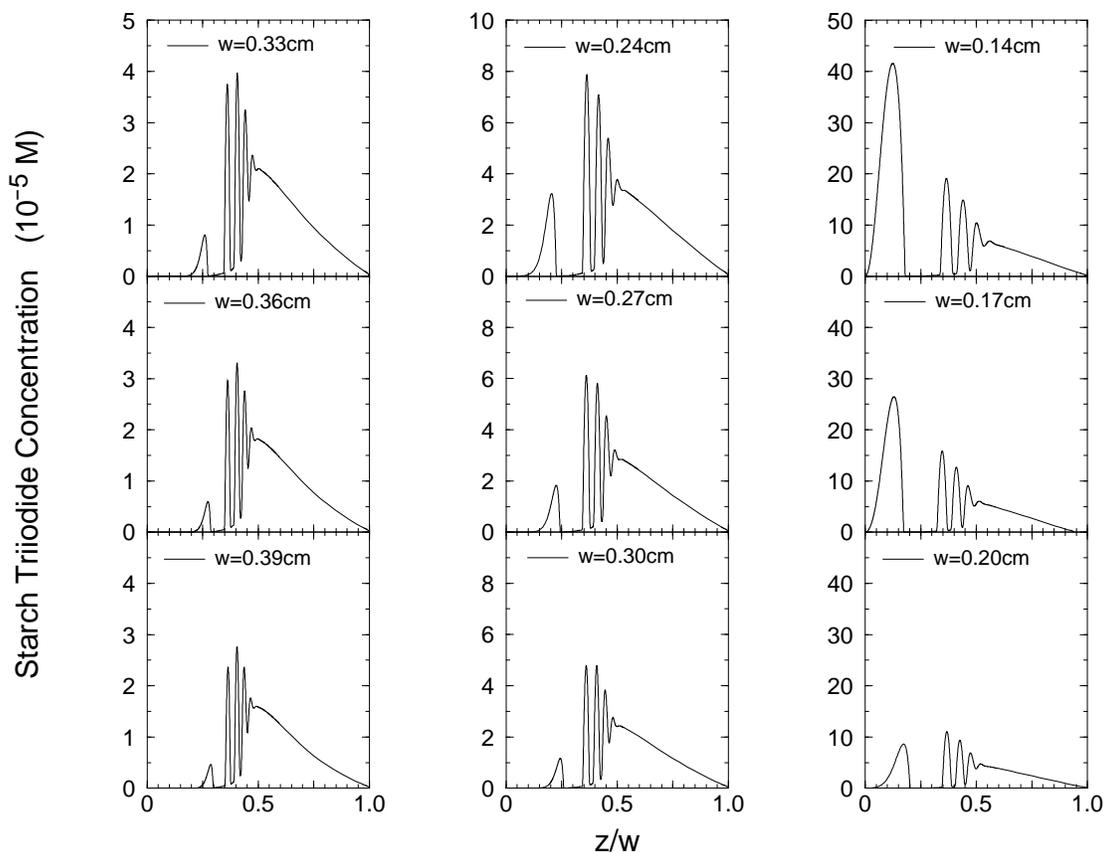}
   \caption[Stationary solution for $\SIIIM$ along feed gradients 
for various gel widths]{\small
Stationary solution for $\SIIIM$ along feed gradients 
for various gel widths: 
This series of plots shows how the stationary state
along the gradients changes as the gel width is varied.
The horizontal axis is the scaled length along the
gradients.  With decreasing gel width, 
we note: (i) shifting of the pattern to the
right, and (ii) increase in the concentration
scale by approximately an order of magnitude 
(primarily due to the leftmost peak).  
}
   \label{fig:fig15}
\end{figure}

\begin{figure}
   \centering \leavevmode
   \epsfxsize=6.0in
   \epsfbox{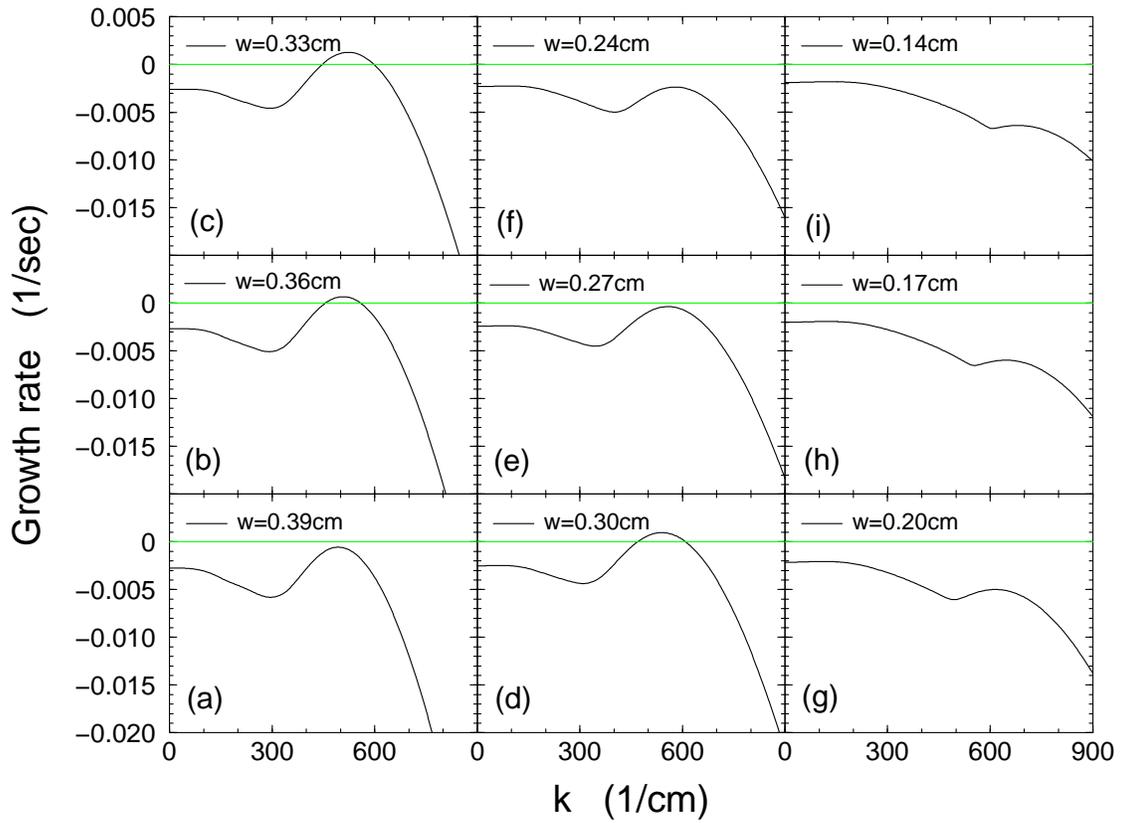}
   \caption[Gain curves corresponding to largest eigenvalue for various
gel widths]{\small
Gain curve corresponding to largest eigenvalue for various
gel widths:  (b)-(d) are unstable; in (g)-(i), the largest
eigenvalue now occurs at $k=0$.
}
   \label{fig:fig16}
\end{figure}

\begin{figure}
   \centering \leavevmode
   \epsfysize=6.0in
   \epsfbox{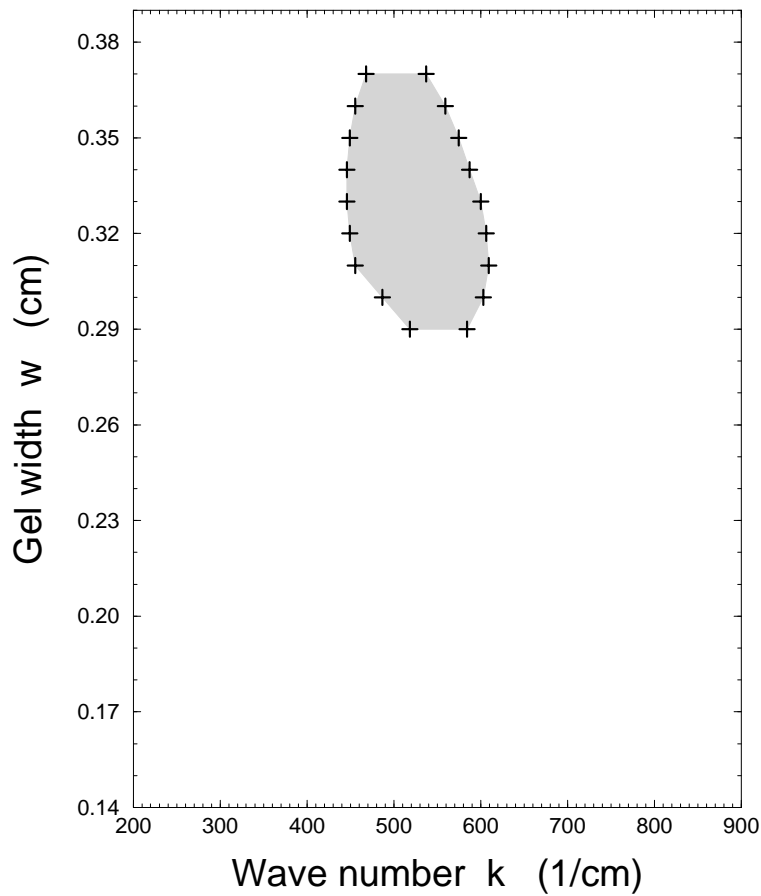}
   \caption[Gel  width $w$ (mm) as a function of 
wave number $k$ of the transverse instability (1/cm)]{\small
Gel  width $w$ (mm) as a function of 
wave number $k$ of the transverse instability (1/cm):  
The crosses represent the marginally stable wave numbers determined from
linear stability analysis of the one-dimensional steady states corresponding
to the given malonic acid feed concentration. A transverse instability
occurs for values of gel width in
the shaded regions, with the range of linearly unstable modes 
delimited by the solid lines for each value of $w$.  The vertical plot range
corresponds to the experimental range of gel width in the bevelled
thin-strip reactor \cite{ref:DDRK_96}.
}
   \label{fig:fig17}
\end{figure}

Our numerical results address the latter question.  
We have reproduced the observed qualitative
trend of the symmetry-breaking instability occurring and subsequently
disappearing in a single layer as the gel width is varied.
The boundary conditions are held fixed at $\MA_{\small L}=0.023$ M,
$\II_{\small R}=0.008$ M and $\CLO_{\small R}=0.006$ M, while the gel width
is varied from $0.14-0.39$ cm.  
The stationary localized patterns along
the gradients as a function of the scaled gel width 
are shown in Figure\ \ref{fig:fig15}. 
Figure\ \ref{fig:fig16} shows the linear stability of each solution
to a symmetry-breaking instability. In Figure\ \ref{fig:fig17},
we have plotted the value of the control parameter $w$
versus wave number $k$, with the solid boundary denoting the
marginally stable wave numbers.  For the parameter values and boundary
conditions numerically investigated here, we do not observe the transition
from an unstable monolayer to an unstable bilayer as the gel width 
is increased.  We have followed the most linearly
unstable $k\neq0$ mode
as the gel width is varied, and it remains singly peaked.

\begin{figure}
   \centering \leavevmode
   \epsfysize=6.0in
   \epsfbox{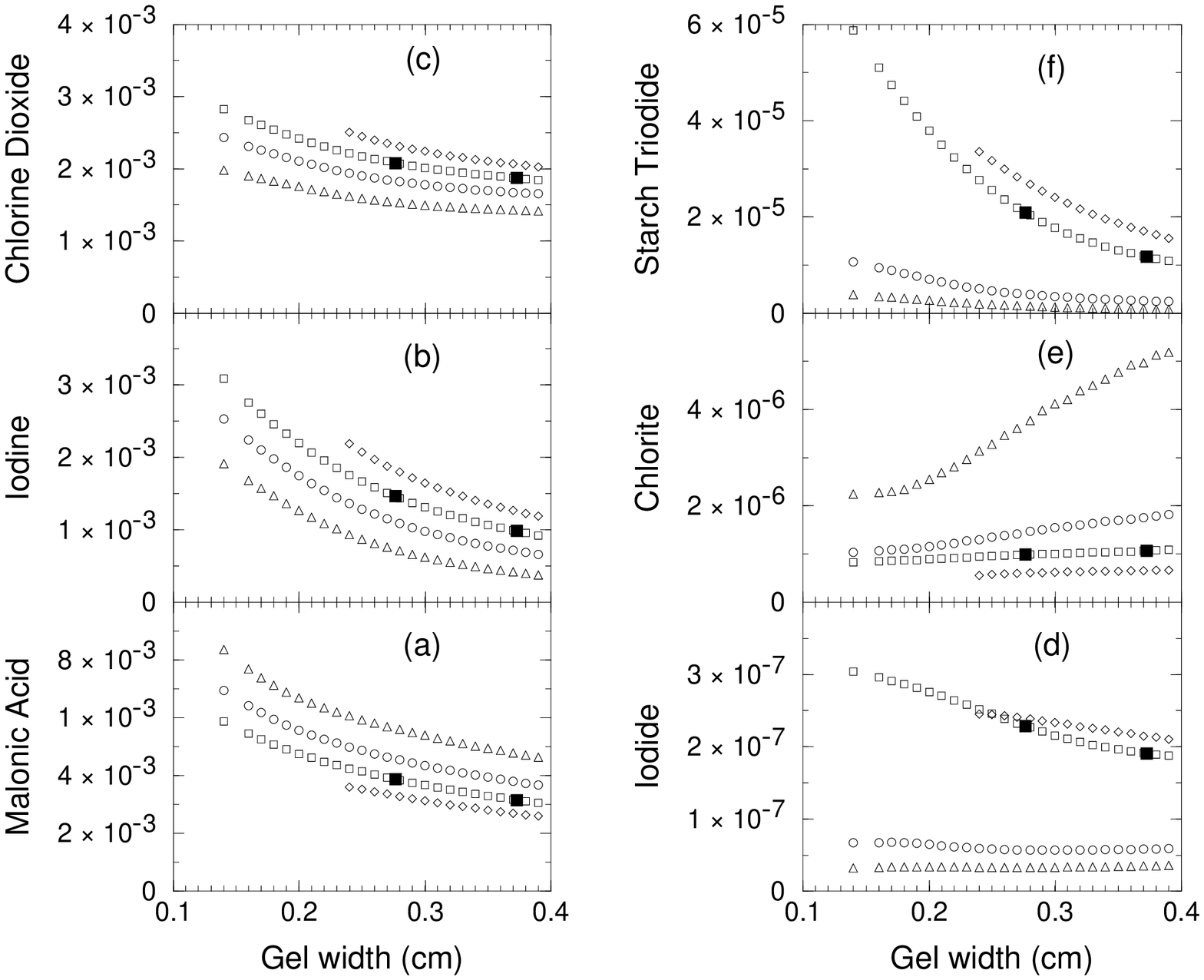}
   \caption[Values of stationary profiles of the CDIMA
chemical species at successive minima of the stationary $\SIIIM$ 
solution as a function of gel width]
{\small
Values of stationary profiles of the CDIMA chemical species at
successive minima of the stationary $\SIIIM$ solution as
a function of gel width, $w$:  The open triangles, circles, squares
and diamonds represent the second, third, fourth and fifth minimum,
respectively.  The filled symbols correspond to the critical
values of gel width; the points between the filled symbols
correspond to the linearly unstable states.  
}
   \label{fig:fig18}
\end{figure}

These numerical results are analogous to those presented 
in Section\ \ref{sec:parsrch_varma},
where the parameter dependence of the Turing instability in a ramped system
was investigated as a function of the malonic acid boundary condition.  Here 
also, we track the concentrations of the chemical species at the location
of the minimum of the starch-triiodide complex where the 
fastest growing/slowest decaying instability eigenvector is localized.  
The results are given in Figure\ \ref{fig:fig18}: 
as before, triangles, circles, squares, and diamonds
represent the second, third, fourth and fifth minimum, respectively.
The solid squares correspond to the critical lengths, below and
above which the instability vanishes.  
We note the following about the 
concentrations of the chemical species in the unstable interval (at the
third minimum of the stationary $\siiim$ solution):
(1)  the concentration of $\clom$ is approximately at $1\times10^{-6}$ M,
in agreement with the variable $\MA_{\small L}$ investigation, 
(2)  the concentration of $\im$ is in the approximate range of $1.8-2.2\times
10^{-7}$ M, again in agreement with 
the variable $\MA_{\small L}$ investigation,  and
(3)  the concentrations of the background species, $\ma$, $\ii$ and $\clo$
are approximately equal to those in the variable $\MA_{\small L}$
case for region III of Figure\ \ref{fig:fig9}.
These results support the simple interpretation that the concentrations of
the dynamical species, the activator $\im$ and the inhibitor $\clom$, are
 key factors in the occurrence of a transverse instability.  
However, this picture may be overly simplified: 
for values of gel width larger than
the upper critical value, the concentrations of $\im$ 
and $\clom$ remain well within
the instability interval of the variable $\MA_{\small L}$ analysis
while the stationary state remains stable.

%%%%%%%%%%%%%%%%%%%%%%%%%%%%%%%%%%%%%%%%%%%%%%%%%%%%%%%%%%%%%%%%%%%%%%%%%%%%%%

\subsubsection{Discussion}
In this section, we have explored the parameter dependence
of the Turing instability as a function of malonic acid boundary condition
and gel width.  
The use of the CIMA reaction-diffusion system 
(as opposed to the CDIMA system) and its 
corresponding boundary species (iodide, rather
than iodine, and chlorite, rather than chlorine) 
by experimenters makes direct
comparison of our numerics with their results difficult. 
Nevertheless, our numerical simulations 
display similar features to those present in experimental results
as  control parameters are varied.
These are: (1) the transition to patterns with progressively
larger numbers of fronts, and 
(2) the appearance and subsequent vanishing of a transverse instability. 
Also agreeing with experimental results, 
the instability is localized near a minimum of the starch
triiodide (as well as iodide) unperturbed solution.
For the parameters and boundary conditions we considered 
in our numerical investigations,
the multiply-peaked linear instability eigenvectors remain stable,
and the experimentally observed transition from a single to
multiple unstable layers is not obtained.
Overall, the agreement between trends in 
these experiments and our numerics is encouraging, and will hopefully
provide motivation for future experiments on the CDIMA system which 
could be compared quantitatively with numerical and analytical results.
 
%%%%%%%%%%%%%%%%%%%%%%%%%%%%%%%%%%%%%%%%%%%%%%%%%%%%%%%%%%%%%%%%%%%%%%%%%%%

\section{Conclusions}

\label{sec:conclude}

In this work, we have focused on the one-dimensional patterns that form in the
presence of feed gradients, a necessary feature of the real experimental
systems. We have shown that longitudinal diffusion along the boundary feed
gradients can be important over length scales longer than the Turing
wavelength. Therefore, the frequently-invoked locally uniform approach for
predicting the linear instability of the stationary patterns along the
gradients to a transverse symmetry-breaking instability is inappropriate in
such cases.

We have also explored the dependence of the Turing instability of these
longitudinal structures on two control parameters. The transition to patterns
with progressively larger number of longitudinal fronts and the appearance and
subsequent vanishing of the transverse instability near a local minimum of the
starch triiodide solution are features which are in agreement with
experimental results. We have attempted to interpret these trends, by
determining that a transverse instability occurs and is localized at that part
of the stationary solution along the gradients where the values of the
concentrations of the dynamical iodide and chlorite species are within a
certain well-defined range.  For the parameters and boundary conditions
investigated here, we do not obtain the experimentally observed transition
from a single to multiple Turing unstable layers. 

Building on the work presented here, one can begin to address many
interesting questions.
At the linear level, it is clear from the numerical solutions that  both the
formation of localized structures along the gradients as well as their
transverse instability are governed by the Turing mechanism. The relationship
between the ``longitudinal'' and ``transverse'' pattern formation is an
interesting question that so far has only been analyzed for model systems
\cite{ref:D_87}. For the realistic chemical description it may be more
feasible to address this question analytically in the limit where the
reservoir concentrations can be approximated by linear profiles: we have found
that these conditions are achieved, for parameter values explored in the
variable gel-width numerical investigation, for small gel widths. A
description of the longitudinal structures can perhaps be sought, in terms of
wave number selection in the presence of control parameter ramps
\cite{ref:MCC_82}. The effect of the variation in background concentration
profiles, with changing gel width or malonic acid reservoir concentration
(Figures\ \ref{fig:fig10} and \ref{fig:fig15}), on these
one-dimensional structures can be investigated within this framework.
Alternatively the symmetry breaking transition might be better understood in
terms of a fully three dimensional periodic instability perturbed by
longitudinal gradients.

At the nonlinear level, the nature of the transition from one-dimensional
non-symmetry breaking front patterns to symmetry-breaking transverse spots in
thin-strip reactors (continuous or discontinuous) can be determined
numerically from the nonlinear evolution of the LRE model in two dimensions
\cite{ref:ssmcc2}. Numerical establishment of the bifurcation behavior for the
two-dimensional ``monolayers'' in disc reactors is also a relevant topic for
further investigation, and would require extending numerical computation to
three dimensions. Dufiet \textsl{et al.} \cite{ref:DB_96} have pointed out
that these monolayers which are confined by a transverse parameter ramp must
be distinguished from what they call \textsl{genuine} two-dimensional
structures which form under uniform control parameters. They have compared
pattern selection in genuine two-dimensional systems and in such monolayers in
the context of an abstract reaction-diffusion model. The advantage of using
the LRE model is that results would then be directly comparable with
experiments based on the CDIMA reaction. An interesting feature of the work
of  Dufiet \textsl{et al. }is the coupling of the pattern forming modes to the
longitudinal displacement of the pattern as a whole: it would be interesting
to investigate this effect using the realistic description of the
longitudinal gradients.

The successful formulation of a realistic model of the chlorine
dioxide-iodine-malonic acid reaction-diffusion system has made this system an
attractive paradigm for the study of nonequilibrium pattern formation
\cite{ref:DBDR_95}. This work represents an attempt at bringing theoretical
and numerical study of pattern formation in chemical systems closer to
experimental studies.

%%%%%%%%%%%%%%%%%%%%%%%%%%%%%%%%%%%%%%%%%%%%%%%%%%%%%%%%%%%%%%%%%%%%%%%%%%%%%%%

\acknowledgments
This work was supported by the National Science Foundation under 
Grant No. DMR-9311444, and by a generous award of computer time from
the San Diego Supercomputer Center. We thank Dan Meiron for useful
advice on the numerics.

%%%%%%%%%%%%%%%%%%%%%%%%%%%%%%%%%%%%%%%%%%%%%%%%%%%%%%%%%%%%%%%%%%%%%%%%%%%%%%%

\appendix

\section{Details of the Turing Instability Conditions}
\label{sec:app_cond}

\subsection{Stability to Homogeneous Perturbations: \mbox{$k=0$}}
\label{sec:ko_pert}
The characteristic equation for $k=0$ is:
  \begin{equation}
        \sigma {\lambda_o}^2 
         - (a_{11} + \sigma a_{22})  \lambda_o
         + (a_{22}a_{11} - a_{12}a_{21}) 
          = 0,
  \end{equation}
        with,
  \begin{equation}
        {\lambda_o}^{(\pm)}  = \frac{1}{2 \sigma} (a_{11} + \sigma a_{22}) 
                     \pm \frac{1}{2 \sigma}\sqrt{(a_{11}+\sigma a_{22})^2 
                             - 4\sigma (a_{11}a_{22}-a_{12}a_{21}) }.
  \end{equation}
We require ${\rm Re} ({\lambda_o}^{(\pm)}) < 0$.
Therefore, in the case of ${\rm Im} ({\lambda_o}^{(\pm)}) \neq 0$, 
we must have:
  \begin{equation}
        \label{eq:c1}
        {\lambda_o}^{(+)} + {\lambda_o}^{(-)} 
           = a_{11} + \sigma a_{22} < 0, 
  \end{equation}
and additionally, in the case of ${\rm Im} ({\lambda_o}^{(\pm)}) = 0$:
  \begin{equation}
        \label{eq:c2}
        {\lambda_o}^{(+)} {\lambda_o}^{(-)} 
           = \sigma \left[ a_{11}a_{22} - a_{12}a_{21} \right] > 0 .
  \end{equation}

%%%%%%%%%%%%%%%%%%%%%%%%%%%%%%%%%%%%%%%%%%%%%%%%%%%%%%%%%%%%%%%%%%%%%%%%%%%%%%%

\subsection{Instability to Inhomogeneous Perturbations: $k \neq 0$}
\label{sec:k_pert}
We require that at least one of the roots be positive for some $k\neq 0$.
Consider the sum of the eigenvalues:
  \begin{equation}
        {\lambda_k}^{(+)} + {\lambda_k}^{(-)} 
           = - (\sigma c + 1) k^2 + (a_{11} + \sigma a_{22}). 
  \end{equation}
Once stability to $k=0$ perturbations is imposed 
according to Eq.\ (\ref{eq:c1}), the above sum will be $<0$, 
and the real part of one of the roots is
necessarily negative.  First, we require that the product of the roots be $<0$
in order to have a positive real root:
  \begin{equation}
        {\lambda_k}^{(+)} {\lambda_k}^{(-)} 
           = \sigma \left( ck^4 - (a_{22} + c a_{11}) k^2
                               + (a_{11}a_{22} - a_{12}a_{21}) \right)
           \equiv h(k^2) < 0.
  \end{equation}
Since the first and third terms in $h(k^2)$ are $ > 0$,  a necessary
condition for $h(k^2) < 0$ is:
  \begin{equation}
        \label{eq:c3}
        a_{22} + c a_{11} > 0 .
  \end{equation}
Furthermore, we know that $h(k^2)$ has a minimum, 
and from Eq.\ (\ref{eq:c2}) that:
\begin{equation}
  h(0) = \sigma (a_{11}a_{22} - a_{12}a_{21}) > 0.
\end{equation}
So, for $h(k^2)<0$, $k^2$ must lie between the two roots ${k_-}^2$ and
${k_+}^2$:
\begin{equation}
\label{eq:k2rng}
  {k_\pm}^2 = \frac{1}{2c} \left[ (a_{22} + ca_{11}) 
               \pm \sqrt{(a_{22}+ca_{11})^2 
                         - 4c(a_{11}a_{22}-a_{12}a_{21})} \right] .
\end{equation}
For ${k_\pm}^2$ to be real, we must have:
\begin{equation}
   \label{eq:c4}
   (a_{22} + c a_{11})^2 - 4c (a_{11}a_{22} - a_{12}a_{21}) \geq 0 .
\end{equation}
Eq.\ (\ref{eq:c4}) is required for real values ${k_\pm}^2$ of the
marginal modes, and Eq.\ (\ref{eq:c3}) is required for ${k_\pm}^2 \ge 0$.
Equations (\ref{eq:c1}),
(\ref{eq:c2}), (\ref{eq:c3}) and (\ref{eq:c4}) constitute the
Turing conditions.

%%%%%%%%%%%%%%%%%%%%%%%%%%%%%%%%%%%%%%%%%%%%%%%%%%%%%%%%%%%%%%%%%%%%%%%%%%%%%%

\subsection{Oscillatory Instability}
\label{sec:osc}
From Eq.\ (\ref{eq:l1}) there will
be a complex conjugate pair of eigenvalues, ${\rm Im} (\lambda_k) \neq 0$,
for: 
\begin{equation}
g(k^2) \equiv (\sigma c -1)^2 k^4 + 2k^2(\sigma c-1)(a_{11} -\sigma a_{22})
          + (a_{11} - \sigma a_{22})^2 + 4\sigma a_{12}a_{21} \le 0.
\end{equation}
To determine the behavior of $g(k^2)$, look at its roots:
\begin{equation}
\label{eq:kh_plus}
  {k_\pm^{({\rm H})}}^2 =  -\frac{(a_{11} - \sigma a_{22}) }{\sigma c - 1}
                   \pm \frac{\sqrt{-4 \sigma a_{12} a_{21}}}{\sigma c - 1}.
\end{equation}
For ${\rm Im} (\lambda_k) \neq 0$, require ${k_+^{({\rm H})}}^2 \ge 0$.
Under typical experimental conditions, we expect $ \sigma c - 1 > 0$. 
We note that $g(k^2)$ possesses a minimum with:
\begin{equation}
\label{eq:hopf}
  g(0) = (a_{11} - \sigma a_{22})^2 + 4\sigma a_{12} a_{21}.
\end{equation}
Therefore, for ${k_+^{({\rm H})}}^2 \ge 0$, we must have $g(0) < 0$, in
which case ${\rm Im}\, \lambda_k \neq 0$ for $0 < k^2 < {k_+^{({\rm H})}}^2$.
The real part of the complex conjugate pair is given by:
\begin{equation}
  {\rm Re}\, \lambda_k =  -\frac{1}{2} \left[ 
              (\sigma c + 1) k^2 - (a_{11} + \sigma a_{22}) \right ],
\end{equation}
and behaves as:
\begin{eqnarray}
  \left. i \right) a_{11} + \sigma a_{22} < 0 
         \quad & \Rightarrow & \quad
               {\rm Re}\, \lambda_k < 0 \\
  \left. ii \right) a_{11} + \sigma a_{22} > 0 
         \quad & \Rightarrow & \quad
               {\rm Re}\, \lambda_k < 0 
                   \quad {\rm for} \quad {k_+^{({\rm H})}}^2 > k^2 > 
                         \frac{a_{11} + \sigma a_{22}}{\sigma c + 1} \\ 
\label{eq:osc_unstab}
         \quad & & \quad
               {\rm Re}\, \lambda_k \geq 0 
                   \quad {\rm for} \quad 0 < k^2 \leq
                         \frac{a_{11} + \sigma a_{22}}{\sigma c + 1}.
\end{eqnarray}

%%%%%%%%%%%%%%%%%%%%%%%%%%%%%%%%%%%%%%%%%%%%%%%%%%%%%%%%%%%%%%%%%%%%%%%%%%%%%%%

%%%%%%%%%%%%%%%%%%%%%%%%%%%%%%%%%%%%%%%%%%%%%%%%%%%%%%%%%%%%%%%%%%%%%%%%%%%%%%%

%%%%%%%%%%%%%%%%%%%%%%%%%%%%%%%%%%%
%         FIGURE  1               %
%%%%%%%%%%%%%%%%%%%%%%%%%%%%%%%%%%%
%
\clearpage

%%%%%%%%%%%%%%%%%%%%%%%%%%%%%%%%%%%
%         FIGURE  2               %
%%%%%%%%%%%%%%%%%%%%%%%%%%%%%%%%%%%
%
\clearpage

%%%%%%%%%%%%%%%%%%%%%%%%%%%%%%%%%%%
%         FIGURE  3               %
%%%%%%%%%%%%%%%%%%%%%%%%%%%%%%%%%%%
%
\clearpage

%%%%%%%%%%%%%%%%%%%%%%%%%%%%%%%%%%%
%         FIGURE  4               %
%%%%%%%%%%%%%%%%%%%%%%%%%%%%%%%%%%%
%
\clearpage

%%%%%%%%%%%%%%%%%%%%%%%%%%%%%%%%%%%
%         FIGURE  5               %
%%%%%%%%%%%%%%%%%%%%%%%%%%%%%%%%%%%
%
\clearpage

%%%%%%%%%%%%%%%%%%%%%%%%%%%%%%%%%%%
%         FIGURE  6               %
%%%%%%%%%%%%%%%%%%%%%%%%%%%%%%%%%%%
%
\clearpage

%%%%%%%%%%%%%%%%%%%%%%%%%%%%%%%%%%%
%         FIGURE  7               %
%%%%%%%%%%%%%%%%%%%%%%%%%%%%%%%%%%%
%
\clearpage

%%%%%%%%%%%%%%%%%%%%%%%%%%%%%%%%%%%
%         FIGURE  8               %
%%%%%%%%%%%%%%%%%%%%%%%%%%%%%%%%%%%
%
\clearpage

%%%%%%%%%%%%%%%%%%%%%%%%%%%%%%%%%%%
%         FIGURE  9               %
%%%%%%%%%%%%%%%%%%%%%%%%%%%%%%%%%%%
%
\clearpage

%%%%%%%%%%%%%%%%%%%%%%%%%%%%%%%%%%%
%         FIGURE 10               %
%%%%%%%%%%%%%%%%%%%%%%%%%%%%%%%%%%%
%
\clearpage

%%%%%%%%%%%%%%%%%%%%%%%%%%%%%%%%%%%
%         FIGURE 11               %
%%%%%%%%%%%%%%%%%%%%%%%%%%%%%%%%%%%
%
\clearpage

%%%%%%%%%%%%%%%%%%%%%%%%%%%%%%%%%%%
%         FIGURE 12               %
%%%%%%%%%%%%%%%%%%%%%%%%%%%%%%%%%%%
%
\clearpage

%%%%%%%%%%%%%%%%%%%%%%%%%%%%%%%%%%%
%         FIGURE 13               %
%%%%%%%%%%%%%%%%%%%%%%%%%%%%%%%%%%%
%
\clearpage

%%%%%%%%%%%%%%%%%%%%%%%%%%%%%%%%%%%
%         FIGURE 14               %
%%%%%%%%%%%%%%%%%%%%%%%%%%%%%%%%%%%
%
\clearpage

%%%%%%%%%%%%%%%%%%%%%%%%%%%%%%%%%%%
%         FIGURE 15               %
%%%%%%%%%%%%%%%%%%%%%%%%%%%%%%%%%%%
%
\clearpage

%%%%%%%%%%%%%%%%%%%%%%%%%%%%%%%%%%%
%         FIGURE 16               %
%%%%%%%%%%%%%%%%%%%%%%%%%%%%%%%%%%%
%
\clearpage

%%%%%%%%%%%%%%%%%%%%%%%%%%%%%%%%%%%
%         FIGURE 17               %
%%%%%%%%%%%%%%%%%%%%%%%%%%%%%%%%%%%
%
\clearpage

%%%%%%%%%%%%%%%%%%%%%%%%%%%%%%%%%%%
%         FIGURE 18               %
%%%%%%%%%%%%%%%%%%%%%%%%%%%%%%%%%%%
%
\clearpage

%%%%%%%%%%%%%%%%%%%%%%%%%%%%%%%%%%%%%%%%%%%%%%%%%%%%%%%%%%%%%%%%%%%%%%%%%%%%%%%

%%%%%%%%%%%%%%%%%%%%%%%%%%%%%%%%%%%
%          TABLE 1                %
%%%%%%%%%%%%%%%%%%%%%%%%%%%%%%%%%%%
%
\clearpage
\begin{table}[hbt]
  \begin{center}
  \caption[Kinetic constants used in the LRE model]{%\small 
             Kinetic constants for the CDIMA system.
            }
  \label{table:T1}
  \begin{tabular}{ccc}
   Rate or diffusion constant & Dimensions  & Value \\
  \hline
   & & \\
   $k_{1a}$  & (s$^{-1}$)          & $9  \times 10^{-4}$ \footnotemark[1]   \\
   $k_{1b}$  & (M)                 & $5  \times 10^{-5}$ \footnotemark[1] \\
   $k_{2}$   & (M$^{-1}$s$^{-1}$)  & $1  \times 10^3$ \footnotemark[1]    \\
   $k_{3a}$  & (M$^{-2}$s$^{-1}$)  & $1.2\times 10^2$ \footnotemark[1]    \\
   $k_{3b}$  & (s$^{-1}$)          & $1.5\times 10^{-4}$ \footnotemark[1] \\
   $h$       & (M$^2$)             & $1.0\times 10^{-14}$ \footnotemark[1] \\
   $k_+$     & (M$^{-2}$s$^{-1}$)    & $6.0\times 10^5$ \footnotemark[2]    \\
   $k_-$  & (s$^{-1}$)    & $1.0$ \footnotemark[2]    \\
   $D_{\im}$ & (cm$^2$s$^{-1}$)      & $7.0\times 10^{-6}$ \footnotemark[3] \\
   $D_{\clom}$ & (cm$^2$s$^{-1}$)    & $7.0\times 10^{-6}$ \footnotemark[3] \\
   $D_{\ii}$ & (cm$^2$s$^{-1}$)      & $6.0\times 10^{-6}$ \footnotemark[1] \\
   $D_{\ma}$ & (cm$^2$s$^{-1}$)      & $4.0\times 10^{-6}$ \footnotemark[1] \\
   $D_{\clo}$ & (cm$^2$s$^{-1}$)     & $7.5\times 10^{-6}$ \footnotemark[1] \\
   $D_{\hp}$ & (cm$^2$s$^{-1}$)     & $1.0\times 10^{-5}$ \\
   $K[\rm S]_o$ & (M$^{-1}$)      & $6.25\times 10^{4}$ \footnotemark[4]\\
  \hline
\multicolumn{3}{l}{
\footnotemark[1]{From \cite{ref:LE_91} at $7^\circ {\rm C}$}; 
\footnotemark[2]{From \cite{ref:KLE_95} at $4^\circ {\rm C}$}; 
\footnotemark[3]{From \cite{ref:LKE_92} at $4^\circ {\rm C}$}; 
\footnotemark[4]{From \cite{ref:LE_95} at $4^\circ {\rm C}$}.
}
  \end{tabular}
  \end{center}
\end{table}
%%%%%%%%%%%%%%%%%%%%%%%%%%%%%%%%%%%%%%%%%%%%%%%%%%%%%%%%%%%%%%%%%%%

%%%%%%%%%%%%%%%%%%%%%%%%%%%%%%%%%%%%%%%%%%%%%%%%%%%%%%%%%%%%%%%%%%%%%%%%%%%%%%%


\begin{references}

 \bibitem{ref:CH_93}    M. C. Cross and P. C. Hohenberg, 
                        Rev.~Mod.~Phys.~{\bf 65}, 854 (1993).

 \bibitem{ref:OuSw_95}  Q.  Ouyang and H. L. Swinney, 
                        {\sl Chemical Waves and Patterns}, 
                        edited by R. Kapral and K. Showalter (Klewer, 1995),
                        p. 269.

 \bibitem{ref:Tur_52}   Turing, A.M., Phil.~Trans.~R.~Soc.~London,
                        Ser.~B {\bf 327}, 37 (1952).

 \bibitem{ref:CDBD_90}  V. Castets, E. Dulos, J. Boissonade, 
                        and P. De Kepper, 
                        Phys.~Rev.~Lett.~{\bf 64}, 2953 (1990).

 \bibitem{ref:DCDB_91}  P. De Kepper, V. Castets, E. Dulos,
                        and J. Boissonade,
                        Physica D {\bf 49}, 161 (1991).

 \bibitem{ref:OuSw_Nat91} Q. Ouyang and H. L. Swinney, 
                        Nature {\bf 352}, 610 (1991).

 \bibitem{ref:Murray}   J. D. Murray, {\sl Mathematical Biology},
                        (Springer-Verlag, Berlin, 1989), Chp.~15.

 \bibitem{ref:LRE_90}   I. Lengyel, G. Rabai, and I. R. Epstein, 
                        J.~Am.~Chem.~Soc.~{\bf 112}, 4606 (1990).

 \bibitem{ref:LRE2_90}  I. Lengyel, G. Rabai, and I. R. Epstein, 
                        J.~Am.~Chem.~Soc.~{\bf 112}, 9104 (1990).

 \bibitem{ref:DBDR_95}  G. Dewel, P. Borckmans, A. De Wit, B. Rudovics, 
                        J. J. Perraud, E. Dulos, J. Boissonade, and 
                        P. De Kepper, Physica A, {\bf 213}, 181 (1995).

 \bibitem{ref:sth}      S. Setayeshgar, {\sl Turing Pattern Formation in
                        the Chlorine Dioxide-Iodine-Malonic Acid 
                        Reaction-Diffusion System}, Ph.D. Thesis, 
                        California Institute of Technology, 1998.

 \bibitem{ref:PB_92}    J. E. Pearson and W. J. Bruno, 
                        Chaos {\bf 2}, 513 (1992).

% new experimental refs go here

 \bibitem{ref:OuSw_91}  Q. Ouyang and H. L. Swinney, Chaos {\bf 1},
                        411 (1991).

 \bibitem{ref:DDRK_96}  E. Dulos, P. Davies, B. Rudovics, and 
                        P. De Kepper, Physica D, {\bf 98}, 53 (1996).

 \bibitem{ref:PADD_92}  J. J. Perraud, K. Agladze, E. Dulos, 
                        and P. De Kepper, 
                        Physica A {\bf 188}, 1 (1992).

  \bibitem{ref:LKE_92}   I. Lengyel, S. Kadar, and I. R. Epstein, 
                        Phys.~Rev.~Lett.~{\bf 69}, 2729 (1992).

 \bibitem{ref:KLE_95}   S. Kadar, I. Lengyel, and I. R. Epstein, 
                        J.~Phys.~Chem.~{\bf 99}, 4504 (1995).

 \bibitem{ref:LE_92}    I. Lengyel and I. R. Epstein, 
                        Proc.~Natl.~Acad.~Sci.~ {\bf 89}, 3977 (1992).

 \bibitem{ref:LE_91}    I. Lengyel and I. R. Epstein, 
                        Science {\bf 251}, 650 (1991).

 \bibitem{ref:LE_95}    I. Lengyel and I. R. Epstein, 
                        {\sl Chemical Waves and Patterns}, 
                        edited by R. Kapral and K. Showalter (Klewer, 1995),
                        p. 297.

 \bibitem{ref:LKE_93}   I. Lengyel, S. Kadar, and I. R. Epstein, 
                        Science {\bf 259}, 493 (1993).

 \bibitem{ref:Num_Recp} W. H. Press, S. A. Teukolsky, W. T. Vetterling, and 
                        B. P. Flannery, {\sl Numerical Recipes in C}
                        (Cambridge Univ.~Press, 1992).

 \bibitem{ref:OS_91}    Q. Ouyang and H. L. Swinney, Chaos~{\bf 1} (4), 
                        411 (1991). 

 \bibitem{ref:eispack}  B. T. Smith, J. M. Boyle, J. J. Dongarra, 
                        B. S. Garbow, Y. Ikebe, V. C. Klema, and C. B. Moler, 
                        {\sl Matrix Eigensystem Routines -- EISPACK Guide}
                        (Springer-Verlag, Berlin, 1976).

 \bibitem{ref:AN_75}    J. F. Auchmuty, and G. Nicolis, 
                        Bulletin of Mathematical Biology {\bf 37}, 
                        323 (1975).


 \bibitem{ref:DB_89}    G. Dewel and P. Borckmans, 
                        Physics Letters A {\bf 138} (4), 189 (1989).

 \bibitem{ref:JPDB_93}  O. Jensen, V. O. Pannbacker, G. Dewel, and 
                        P. Borckmans, Physics Letters A {\bf 179}, 91 (1993).
 
 \bibitem{ref:JPMDB_93} O. Jensen, V. O. Pannbacker, 
                        E. Mosekilde, G. Dewel, and P. Borckmans, 
                        Phys.~Rev.~E {\bf 50}, 736 (1993). 
                      
 \bibitem{ref:ssmcc2}   S. Setayeshgar and M. C. Cross, article in 
                        preparation.  

 \bibitem{ref:D_87}     G. Dewel, D. Walgraef, and P. Borckmans, 
                        J.~Chim.~Phys.~(France) {\bf 84}, 1335 (1987).

 \bibitem{ref:MCC_82}   L. Kramer, E. Benjacob, H. Brand, and M. C. Cross,
                        Phys.~Rev.~Lett. {\bf 49}, 1891 (1982).

 \bibitem{ref:DB_96}    V. Dufiet and J. Boissonade, Phys. Rev. E.,
                        {\bf 53}, 4883 (1996).

\end{references}
\end{document}